\documentclass[twocolumn]{emulateapj}
\usepackage{graphicx,subfigure,color}
\pdfoutput=1
\shorttitle{Interferometric imaging of AGN with JWST}
\shortauthors{Ford et al.}

\begin{document}

\newcommand \Dlam     {D_{\lambda}}
\newcommand \dlam     {d_{\lambda}}
\newcommand \blam     {b_{\lambda}}
\newcommand \ssgn     {\,\, ^2\rm sgn}            
\newcommand \shah     {{\rm III}}             
\newcommand \sshah     {{\,\, ^2\rm III}}            
\newcommand \ppi     {\,\, ^2\Pi}                
\newcommand \sinc     {\,\, {\rm sinc}}         
\newcommand \ssinc     {\,\, ^2 {\rm sinc}}        
\newcommand \bx     {{\bf x}}
\newcommand \bk     {{\bf k}}
\newcommand \bkzero {{\bf k_o}}
\newcommand \bu     {{\bf u}}
\newcommand \cf {cf.}
\newcommand \eg {{\it e.g., }}
\newcommand \etc {{\it etc}}
\newcommand \etal {{\it et~al.}}
\newcommand \ie {{\it i.e.,}}
\newcommand \viz {{\it viz.,}}
\newcommand \lap {$\stackrel{<}{\sim}$}
\newcommand \gap {$\stackrel{>}{\sim}$}

\title{AGN and quasar science with aperture masking interferometry \\
		on the James Webb Space Telescope}

\author{K.E. Saavik Ford\altaffilmark{1,2}, Barry McKernan\altaffilmark{1,2}}
	\affil{Department of Science, Borough of Manhattan Community College,
		City University of New York, New York, NY 10007}
\author{ Anand Sivaramakrishnan\altaffilmark{1,3}, Andr\'{e} R.  Martel, Anton Koekemoer,}
	\affil{Space Telescope Science Institute, 3700 San Martin Drive, Baltimore, MD 21218}
\author {David Lafreni\`{e}re}
	\affil{Universit\'e de Montr\'eal,  D\'epartement de Physique \\
		C.P. 6128 Succ. Centre-ville, Qc, H3C 3J7, Canada}
	\and
\author {S\'{e}bastien Parmentier,}
	\affil{Department of Physics and Astronomy, Stony Brook University,
		Stony Brook, NY 11794}

\altaffiltext{1}{Department of Astrophysics, American Museum of Natural History, New York, NY 10024}
\altaffiltext{2}{CUNY Graduate Center, 365 5th Avenue, New York, NY 10016}
\altaffiltext{3}{Department of Physics \& Astronomy, Stony Brook University, Stony Brook, NY 11794}

\begin{abstract}
Due to feedback from accretion onto supermassive black holes (SMBHs),
Active Galactic Nuclei (AGNs) are believed to play a key role in
$\Lambda$CDM cosmology and galaxy formation.
However, AGNs' extreme luminosities and
the small angular size of their accretion flows
create a challenging imaging problem.
We show James Webb Space Telescope's Near Infrared Imager and Slitless
Spectrograph (JWST-NIRISS) Aperture Masking Interferometry (AMI) mode
will enable true imaging
(\ie\ without any requirement of prior assumptions on source geometry)
at $\sim$~$65$~mas angular resolution at the centers of AGNs.
This is advantageous for studying complex extended accretion flows around SMBHs,
and in other areas of angular-resolution-limited astrophysics.
By simulating data sequences incorporating expected sources of noise,
we demonstrate that JWST-NIRISS AMI mode can map extended structure
at a pixel-to-pixel contrast of $\sim$~$10^{-2}$ around an L=$7.5$ point source, using short exposure times (minutes).
Such images will test models of AGN feedback, fuelling and
structure (complementary with ALMA observations), and are not currently 
supported by any ground-based
IR interferometer or telescope. 
Binary point source contrast with NIRISS is $\sim$~$10^{-4}$
(for observing binary nuclei in merging galaxies),
significantly better than current ground-based
optical or IR interferometry.
JWST-NIRISS' seven-hole non-redundant mask has a throughput of 15\%, and
utilizes NIRISS' F277W (2.77\micron), F380M (3.8\micron), F430M (4.3\micron),
and F480M (4.8\micron) filters.
NIRISS' square pixels are 65~mas per side, with a field of view
$\sim$~$2\arcmin$~$\times$~$2\arcmin$.
We also extrapolate our results to AGN science enabled by non-redundant masking on future 2.4~m and 16~m 
space telescopes working at long-UV to near-IR wavelengths.
\end{abstract}


\section{Introduction} \label{sec:intro} Supermassive black holes (SMBHs),
with masses $\sim 10^{6}-10^{9}M_{\odot}$ are believed to lie in the centers
of nearly all galaxies in the Universe \citep[e.g.][]{kr95}. When accreting,
these SMBHs can outshine their host galaxy, with
luminosities spanning $\sim 10^{10}-10^{16}L_{\odot}$ \citep[e.g.][]{ho08,kaufheck09}.
The resulting active galactic nuclei (AGNs) and quasars are believed to play a
key role in galaxy formation and, via feedback, in $\Lambda$CDM cosmology
\citep[e.g.][]{springel06,schawinski07,silknusser10}. In spite of their importance, the combination of
extreme luminosity ($\sim 10^{10}-10^{16}L_{\odot}$), small size ($\sim$pc) and
large distance ($\geq 10$~Mpc typically) makes it very difficult to image
details in most AGNs. Spectral and variability studies have allowed us to infer
a great deal about AGNs. However high contrast and high resolution images would
provide new, strong tests of models of AGN binarity, structure, fuelling and feedback,
which in turn impacts models of galaxy formation and $\Lambda$CDM cosmology. 

A non-redundant mask (NRM) placed in a pupil plane of a telescope converts a
traditional single aperture telescope into an interferometer
\citep[e.g.][]{bald86,readhead88,tut98,mon03,siva09}.  This enables moderate contrast, high
resolution imaging. This method has long been used with ground-based optical
and near-IR telescopes to study faint structure and point sources around stars
in our own Galaxy \citep[e.g.][]{han87,tut00,mon03}.
However, ground-based optical and IR interferometers do not
measure fringe phase reliably.  Instead particular combinations of fringe
phases known as \emph{closure phases} are used to fit models selected \emph{a
priori}, because these closure phases are calibratable observables.
Calibration of closure phases is accomplished by observing a known point
source.  In the cores of AGNs there are a wide range of expected structures, so
\emph{true imaging} provides more science than model fitting.
However, due to atmospheric turbulence, ground-based model fitting of optical and IR
interferometric closure quantities does not allow us to construct morphologies
reliably \citep[see e.g.][and \S\ref{sec:agn}, \S\ref{sec:groundcomp}, and \S\ref{sec:sample}, especially \S\ref{sec:n4151}]{RTM11,bar12}.

The James Webb Space Telescope (JWST) Near Infrared Imager and Slitless
Spectrograph\footnote{NIRISS, built by COM DEV Canada, is a Canadian Space Agency
contribution to JWST.} \citep{doyon12} will deploy a non-redundant mask (NRM) in
space \citep{siva12}.  Above the atmosphere fringe phase and amplitude should be
easily measured in the optical and near-IR, enabling model-free imaging at a
resolution of $\lambda/2D$ ($\lambda$ being the wavelength, and $D$ the
telescope diameter or longest baseline length).  Since space-based observing
also provides both closure phases and closure amplitudes, those targets
possessing known priors can utilize these closure quantities to improve
estimates of model parameters. Good closure amplitudes improve model fits,
especially those that include symmetric structures
\citep[e.g.][]{readhead80}.  On the ground closure amplitudes have been of
questionable utility in many cases, because changing atmospheric conditions
make calibration very difficult.
In this paper we demonstrate the potential of space-based NRM for AGN and
quasar science.  We focus on JWST-NIRISS' Aperture Masking Interferometry mode (AMI), to show that it should, for the
first time ever,  be able to image details of the accretion flow in AGNs.
We extrapolate our near-IR predictions to aperture masking on possible
future optical/UV space telescopes to outline their potential for SMBH science.

In \S\ref{sec:agn} we discuss how images of the extended accretion flow could
help answer outstanding scientific questions concerning AGNs and quasars. In
\S\ref{sec:nrm} using NIRISS in AMI mode as an example, we show how a space telescope
can operate as an imaging interferometer, and make comparisons to ground-based
instruments.  In \S\ref{sec:sample} we discuss what NIRISS in AMI mode could observe in
some representative celestial objects.
In \S\ref{sec:future} we briefly discuss NRM on possible future missions.

\section{The need for imaging}
\label{sec:agn}

AGN models are strongly constrained by spectral and variability studies, but
not yet by imaging studies (apart from prominent jets).  A
(model fitted) dusty torus has been imaged in NGC 1068 due to uniquely
favorable conditions--it is a very nearby Seyfert 2, and is therefore highly
obscured by the torus, while still being bright enough to be seen using
interferometric techniques, without requiring significant contrast \citep{raban09}.
Images of AGNs at both moderate to high contrast \emph{and} high resolution
will show the morphology of structures feeding black hole accretion, and
feedback on the host galaxy. Both are areas where geometry could rule out models.
We outline several AGN science goals that require optical/infrared (O/IR) imaging,
both those achievable with NIRISS in AMI mode and those that will require future
instruments.

\subsection{Dual and Binary AGNs}
\label{sec:binaries}

The standard cosmological model of hierarchical galactic mergers should produce
large numbers of merging SMBHs. If both SMBHs are
accreting, we should observe large numbers of dual AGN in galaxies. However,
only a handful of dual AGN have been observed to date
\citep[e.g.][]{liu10,comerford12}. At optical wavelengths, dual AGN candidates are
selected from nuclei displaying double-peaked narrow emission lines with
velocity offsets relative to stellar absorption lines
\citep[e.g.][]{comerford13}. Follow-up in the near IR and optical can
reveal tidal features and separated stellar populations \citep{liu10}. The
projected spatial offsets between dual AGNs in SDSS selected samples are $\sim
1\arcsec$, corresponding to $\sim$kpc separations \citep{comerford12}. With an inner working angle (IWA)
of $\sim 70$~mas, NIRISS in AMI mode will be capable of searching for dual AGNs in the near IR
at angular separations at least an order of magnitude smaller than presently
achieveable. A NIRISS AMI mode survey of low redshift galactic nuclei displaying double-peaked
optical emission lines will find dual AGNs down to separations
of $ \leq 17$~pc and contrasts of $\leq 10$~mag in galaxies out to
$\sim 50$~Mpc. Thus, NIRISS' AMI will probe binaries closer to merger and with lower accretion rate (and
lower mass) secondary SMBHs. In addition, space-based NRM (unlike ground-based
NRM--see \S\ref{sec:groundcomp}) will yield astrometric information on the galactic nucleus, allowing us to
distinguish AGNs offset from the dynamical center of the galaxy. Astrometry in
these cases allows us to test models of SMBH recoil and SMBH mergers where only
the secondary is accreting.

\subsection{How are AGNs and quasars fuelled?} 
\label{sec:fuelling}
For SMBHs to feed at the inferred fractions of the Eddington 
rate \citep{kaufheck09}, large masses of gas must 
lose a large amount of angular momentum. At present 
we do not understand whether the process of angular momentum loss is
continuous or a sudden, single, violent event. Nuclear bars, nuclear spirals or 
rings of star formation might provide a continuous supply of low angular momentum 
gas over a long time \citep[e.g.][]{hop10b,schartmann10}. Or, fuel could be delivered by 
one-time events such as the infall of giant molecular clouds \citep{hh06} or 
cloud bombardment from the halo \citep{whc10}.  
The maximum energy extractable from the mass reservoirs that fuel activity is $\sim \eta M c^{2} \sim 4 \times 10^{59}(10^{61})$~ergs respectively, where $\eta$ is the accretion efficiency ($\eta  \sim 0.06(0.42)$ for a static (maximally spinning) black hole. Assuming $\eta \sim 0.1$ if a 
quasar is generated by accretion at the Eddington 
rate onto a $10^{8}M_{\odot}$ SMBH, then a quasar lifetime of 1(100)~Myr requires a mass reservoir of $M \sim 2\times 10^{6}(10^{8}M_{\odot})$ respectively. Therefore, powering a quasar for $\sim 1$~Myrs requires fuelling by a mass equivalent to a giant molecular cloud. But the mass equivalent of an entire dwarf galaxy (probably in a minor merger) would be required to fuel the quasar for $100$~Myr. Evidently, if we can 
image fuelling structures around quasars we can constrain: the size of the mass 
reservoir, the lifetime and fuelling mechanisms. Many of these targets will be challenging for JWST. However,
for 3C 273 (a $10^{9}M_{\odot}$ SMBH) to accrete at the Eddington rate 
($\sim 20M_{\odot}$/yr) for only 10~Myrs, requires $\sim 2 \times 10^{8}M_{\odot}$ of fuel. 
Such a reservoir should be visible in moderate to high contrast images of the 
regions around quasars $\sim100$'s pc from the core (potentially observable by NIRISS in AMI mode).
Beyond simple detection, images revealing the geometry of a fuelling structure will allow us to distinguish 
between models of continuous nuclear fuelling by a nuclear bar or spiral 
on one hand, and intermittent supply mechanisms on the other.
Additionally, if no large reservoir is found we must consider models of quasar luminosity due to processes solely internal 
to the AGN accretion disk, for instance the migration of stars or compact objects within the quasar \citep{gr01,mck11b}.

\subsection{Constraining $\Lambda$CDM cosmology and theories of galaxy formation}
\label{sec:lcdm}
Fuelling also directly impacts feedback--a key ingredient in
$\Lambda$CDM cosmology--since the energy available for feedback is limited
by the energy available through fuelling, and feedback can only occur while
the AGN is active.  The effects of feedback are also
affected by geometry; hence imaging is an important tool for investigating feedback.
We infer large mass outflows in many 
AGNs \citep{mck07,miniutti10}. Such outflows can have a large impact on the host 
galaxy, suppressing star formation at late cosmological times \citep{springel06,sij07,silknusser10}. Such suppression may be required
to create galaxies as we see them today, possibly including the observed $M-\sigma$ relation 
between SMBH mass and stellar velocity dispersion in galactic bulges \citep{fm00}. 
The standard model of $\Lambda$CDM cosmology predicts hierarchical growth of structures, but may require significant AGN 
feedback in order to explain present-day structure in our Universe 
\citep[e.g.][]{springel05,bower06,schawinski07,mccarthy10}. 

The impact of feedback from a quasar on its host galaxy depends on the total energy 
output, but also its duration and geometry. Constraining quasar lifetimes (as in \S\ref{sec:fuelling}) itself constrains
feedback. If quasars are short-lived \citep{martini03} we must 
re-think the role of AGN feedback in both galaxy formation and $\Lambda$CDM cosmology. Images
will also allow us to determine the geometry of feedback. Broad versus collimated outflows, for example, can have
very different effects on a host galaxy.

Nearby Seyfert galaxies are thought to be qualitatively similar to more distant quasars;
fuelling, lifetimes and feedback remain key questions for these targets as well. If they are
scaled down versions of quasars, high angular resolution, moderate contrast images of
the structure of their fuelling regions (at few-10's~pc, well within the capability of NIRISS in AMI mode) can provide us 
with better models for their distant, brighter analogs.

\subsection{Is the outer accretion disk stable?} 
\label{sec:outer_disk}
Around SMBHs accretion disks should be unstable 
to gravitational collapse beyond $\sim 10^{3}$ gravitational 
radii ($r_{g}=2GM/c^{2}$). The instability 
criterion for gas disks is written in terms of Toomre's Q parameter where
\begin{equation}
Q=\frac{\kappa c_{\rm s}}{\pi G \Sigma}
\label{eq:toomre}
\end{equation}
where $\kappa$ is the gas epicyclic frequency, 
$c_{\rm s}$ the gas sound speed and $\Sigma$ the disk surface density. A gas disk is unstable when $Q<1$.
Since $\kappa c_{\rm s}$ drops more rapidly than $\Sigma$ in the outskirts of realistic gas disk models 
\citep[e.g.][]{sirko03}, we expect outer AGN disks to 
be vulnerable to clumping and rapid collapse into star 
forming regions \citep[e.g.][]{shlosman87,gt04}. The AGN outer disk may therefore consist of many 
unstable, colliding clouds or density fluctuations \citep[e.g.][]{nenk08,mor09}. By imaging the 
outskirts of AGNs we can put limits on the number, size, extent, supply and luminosity of warm 
clouds in the outer AGN disk, directly testing the link between star formation and AGNs. 
NIRISS in AMI mode should image the outer accretion disk (torus) of about $20$ nearby AGNs;
future missions will be required to assemble a larger ($\sim100$'s) sample.

\subsection{Searching for gaps and cavities in disks}
\label{sec:gaps}
Analogous to protoplanetary accretion disks, gaps and cavities 
can appear in AGN disks because of the presence of intermediate mass or SMBHs in the AGN disk \citep{arty93,syer95,ivanov99,levin07,koc12,mck14}. The critical mass ratio ($q=q_{\rm crit}$) of secondary black hole to primary SMBH, above which a gap is opened in a disk is \citep{lp86,mck14}
\begin{equation}
q_{\rm crit} \approx \left(\frac{27\pi}{8}\right)^{1/2}\left(\frac{H}{r}\right)^{5/2}\alpha^{1/2}
\end{equation}
where $H/r$ is the disk thickness and $\alpha$ is the viscosity parameter \citep{ss73}.
 Images of gaps and cavities in AGN disks will directly constrain models of disk thickness and viscosity as well as models of intermediate mass black hole formation \citep{mck12a}. Imaging gaps or cavities in AGN disks requires much
higher angular resolution than required for the tests outlined above, but may be possible with future telescopes.

\subsection{Testing the standard model of AGNs} 
\label{sec:stdmodel}
The standard model of AGNs assumes that orientation is a key determinant 
of observed properties \citep{ant93,urpad95}. However, random accretion events 
onto SMBHs should result in misalignment between axisymmetric galactic disks,
AGN accretion disks and the spin of the SMBH \citep{vol07}. 
A rotating black hole will torque the accretion disk into its equatorial plane, leading to a 
warp in a misaligned accretion disk \citep{bp75,kp06}. Thus, obscuration in many AGNs 
could actually be due to misaligned or warped disks \citep{le10}. Images of 
the outskirts of AGN disks will reveal warps and misalignments between the disk and the 
plane of the host galaxy and test models of stochastic accretion.  Alignment checks 
between host galaxy and fuelling region
are achievable with NIRISS in AMI mode for tens of nearby AGNs; 
warping in the outer accretion disk is also
observable for a smaller sample. 
Observing warps in the inner disk awaits instruments with smaller inner working angles.

Another important source of AGN obscuration is a large covering of orbiting clouds, 
independent of the AGN disk orientation \citep[e.g.][]{krolik,weedmanhouck09,mck10a}. The orbiting 
clouds may originate among the unstable clumps believed to exist in the outskirts of the disk.
Images of the outskirts of AGN disks (well within range of NIRISS in AMI mode) will constrain the geometry and
ratio of orbital cloud mass to torus cloud mass and will allow us to test models of the 
origin of the orbiting clouds/stars \citep[e.g.][and references therein]{mck98,risaliti02,turnermiller09}. 

\section{Imaging Extended Structures using JWST-NIRISS AMI}
\label{sec:nrm}
Here we describe the JWST-NIRISS AMI mode, our basic observational strategy
and image reconstruction technique. To make our exposition accessible to those unfamiliar with imaging interferometry, we 
include 3 didactic appendices on basic topics. 
We begin with an explanation of the JWST-NIRISS AMI system; then discuss our simulated observations including
sources of noise; we motivate and describe a move to the Fourier plane; we then explain our standard deconvolution strategy,
showing results for several models. We then demonstrate some extensions to our standard method, and end with a comparison
to ground-based capabilities. We highlight the importance of model-free image 
reconstruction with NIRISS in AMI mode, since \emph{a priori} models of AGNs environs are extremely diverse.

\subsection{JWST-NIRISS AMI implementation}
\begin{figure}
\centerline{
\includegraphics[angle=0,width=7.0cm]{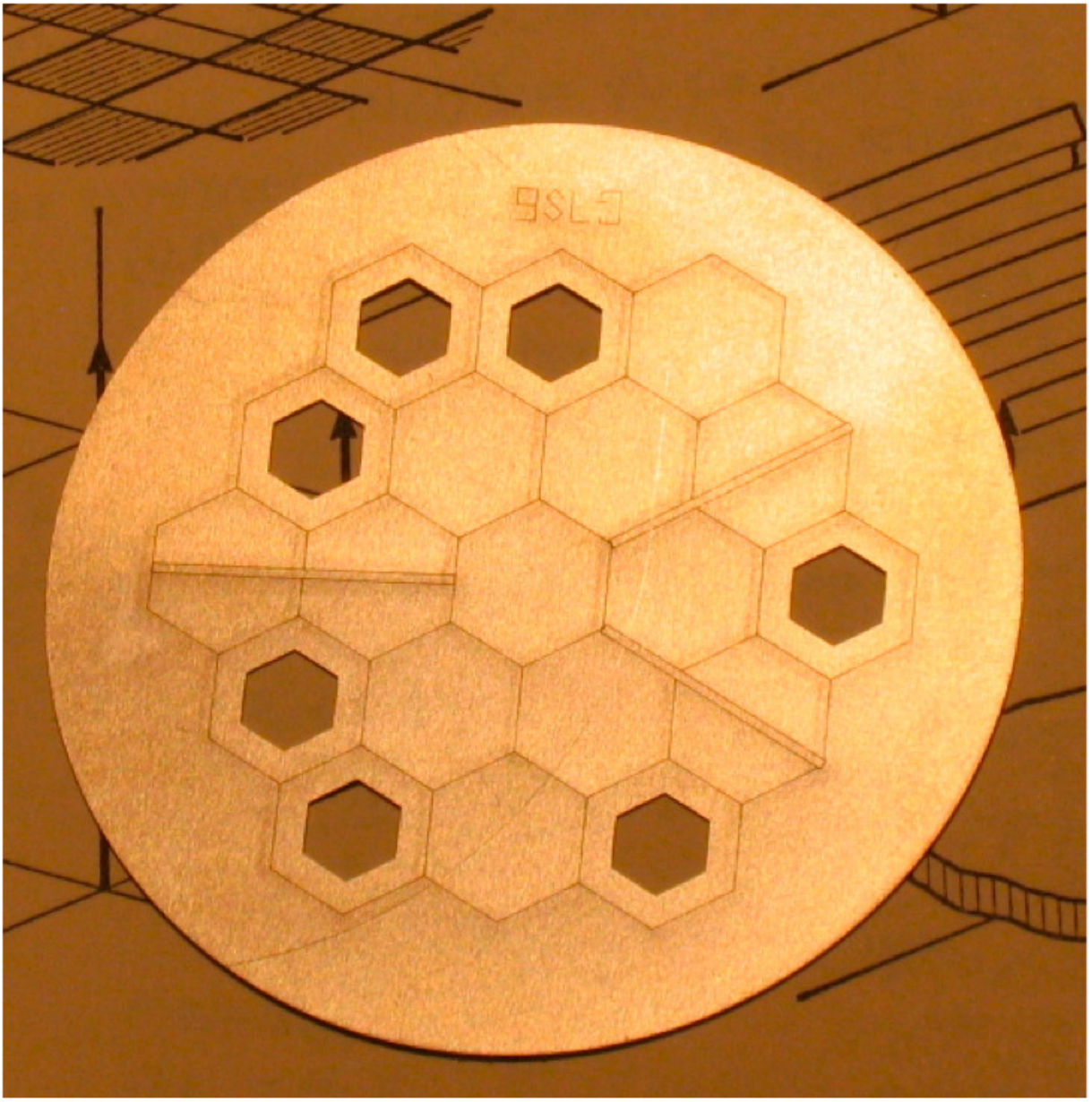}}
\caption{A prototype of the NIRISS 7-hole non-redundant pupil mask (AMI mode) 
with engraved outline of primary mirror segments and secondary supports. Engraving
is on the reverse side of the mask, so this image is looking through the back of it
(flip and rotate to compare to figure~\ref{fig:unmasked}).
The mask has undersized holes compared to 
the segment size to mitigate against pupil shear up to 3.8$\%$. The 
projection of the JWST pupil is nominally circumscribed by a circle 39~mm in 
diameter.  The part diameter is 50~mm.}
\label{fig:mask7fig}
\end{figure}

\begin{figure*}
\centerline{
\includegraphics[angle=0,width=18.0cm]{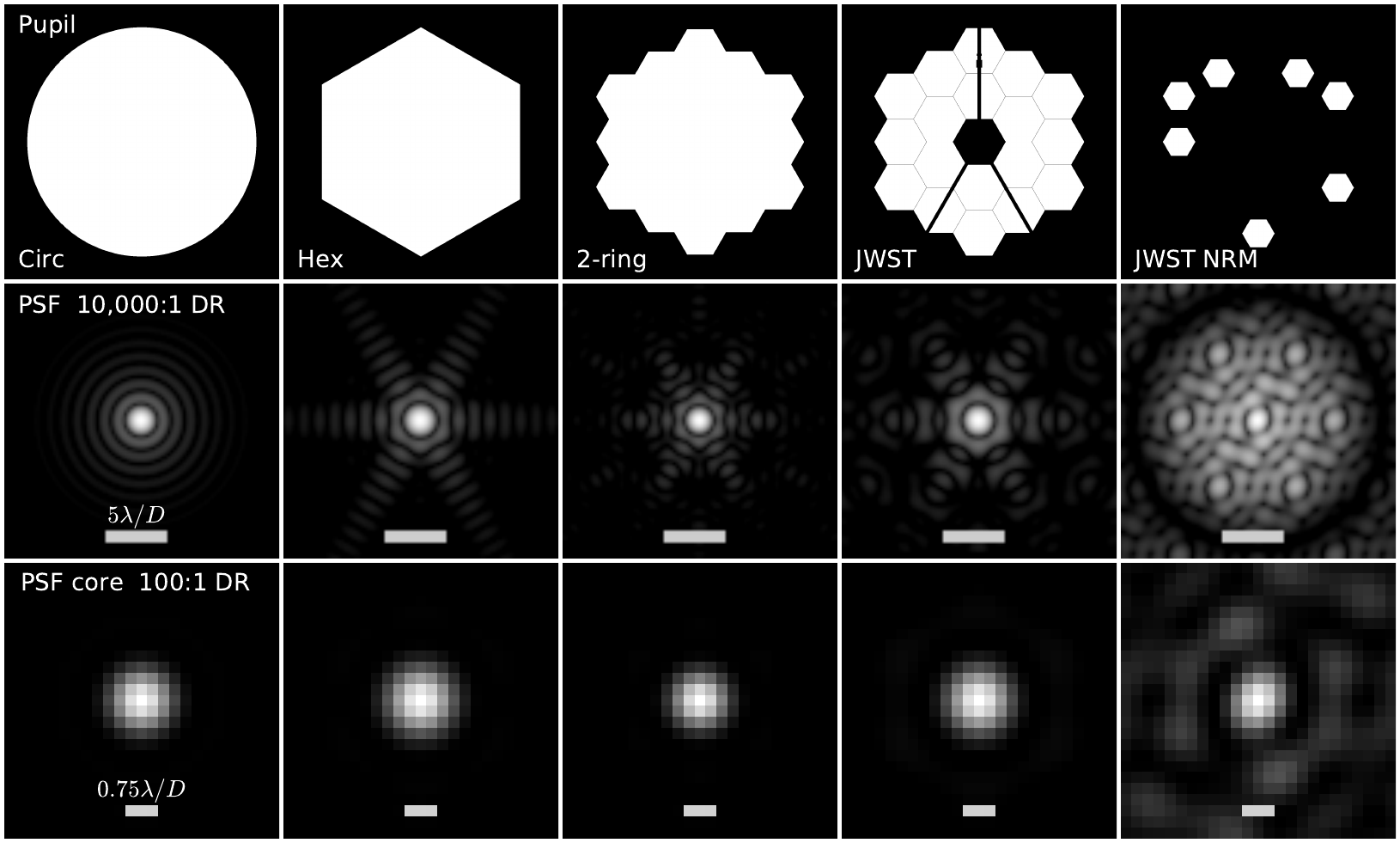}}
\caption{A comparison of the point spread functions of different shaped 
apertures. The top row shows apertures as labelled. 
The middle row shows the corresponding PSFs with 10,000:1 dynamic range, at consistent stretch and pixel scale.
The bottom row shows the cores of the same PSFs with 100:1 dynamic range. Note the relatively small core,
and deep null, surrounding the JWST NRM PSF.
\label{fig:unmasked}}
\end{figure*}

\begin{figure}
\centerline{
\includegraphics[angle=0,width=8.0cm]{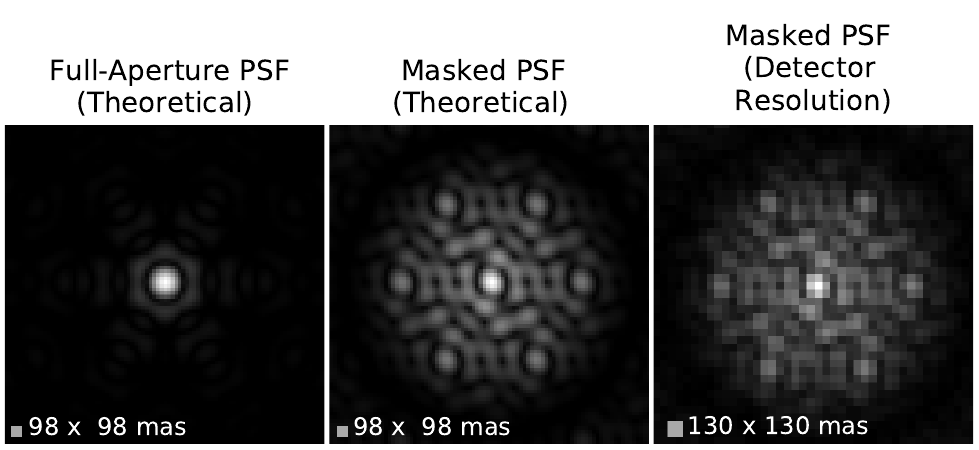}}
\caption{The simulated point spread function for JWST NIRISS, displayed at 
the same dynamic range.  Panels a, b, c from left to right:
a) Oversampled by 2x, no noise, generated from a simulated observation 
of a 7.5~mag point source at 4.5$\micron$.
b) As for a), but with the addition of NRM.
c) Similar to b), but sampled at the resolution of the NIRISS detector (65~mas
per pixel--undersampled given our theoretical resolution).  
Panel c) also includes realistic flat-field noise, photon noise
read noise, intra-pixel sensitivity variations and telescope pointing error, 
for a 0.66 (0.42)~sec integration in F380M, F430M (F480M) filters. Pointing error is (just) detectable by eye.
See text for details of noise model and extrapolation to various
filters.
\label{fig:psfcompare}}
\end{figure}

NIRISS's non-redundant mask consists of a seven-hole, $\sim 15$\% throughput,
titanium mask in a slot in the pupil wheel of NIRISS \citep{doyon12}.  A full-scale
prototype is shown in fig.~\ref{fig:mask7fig}.  The NRM can be used with
NIRISS's F277W, F380M, F430M or F480M filters, which are centered at $2.77,
3.80, 4.30$ and $4.80\micron$, with bandwidths of 25, 5, 5 and 8\%,
respectively. Fig.~\ref{fig:unmasked} demonstrates the effect of increasingly
complex apertures on PSF shape, ending with AMI mode's theoretical PSF. Clearly, it
is more complicated than the unmasked JWST PSF.  However, the core of the NRM
PSF is noticeably smaller, and it is surrounded by a deep null. The theoretical
angular resolution is thus $0.5\lambda/D$ (65~mas at $4\micron$), which is more
than a factor of 2 smaller than the $1.22\lambda/D$ Rayleigh limit.  NIRISS's
65~mas detector pixels are Nyquist-spaced at a wavelength of $4~\micron$.
Fig.~\ref{fig:psfcompare} shows both oversampled and detector pixel-sampled
PSFs  (all simulated sources of noise are included in the last panel).  The
NRM image is most useful in a region between an inner and outer working
angle (IWA and OWA).  For NIRISS, the IWA is $0.5 \lambda/ D$, where $\lambda$ is
the wavelength and $D$ the primary mirror diameter or longest baseline. The OWA
is 4--5~$\lambda/ D$ --- at wider angular scales direct imaging is more
efficient than NRM.  Thus NIRISS' AMI mode is most interesting at separations
between about 65~mas and 600~mas.  Further work may show that in the shortest
wavelength filter, F277W, AMI observations could be dithered to provide slightly
smaller IWA than 65~mas \citep{koe05,greenbaum13}.

\subsection{Simulated observations and noise sources}
\label{sec:pipe}
We simulated observations of a 7.5~magnitude (mag) point source with NIRISS in AMI mode, 
conservatively assuming a detector with a $1\%$ bandpass centered on 4.5$\micron$. 
Exposure times to actual NIRISS filters centered on $3.80, 4.30,$ and $4.80\micron$ are
extrapolated from our initial simulations; what we simulated as a 1~second exposure would actually take 0.25~seconds
with NIRISS F380M or F430M and 0.16~seconds using NIRISS F480M, given their 5, 5 and 8 \% bandwidths, respectively.
The observation consisted of 10 sequences of $N_{exp}=28$ exposures, each exposure equivalent to 0.66 (0.42)~seconds 
for the F380M, F430M and (F480M) filters, respectively.  
Simulated noise sources included: 
a polychromatic PSF according to a specified transmission profile; pixel flat-field error of $0.1\%$; 
variable intra-pixel response with a Gaussian profile (1 at center, $0.8\pm 0.05$(rms) at the corner); 
read noise; dark current; background and photon noise; a pointing error of 5~mas rms per axis assumed between each exposure. 
We expect spectral smearing 
($\sim\lambda /\Delta \lambda$) to be negligible when working within $2.5 \lambda / D$. 

\begin{figure}
\centerline{
\includegraphics[angle=0,width=8.0cm]{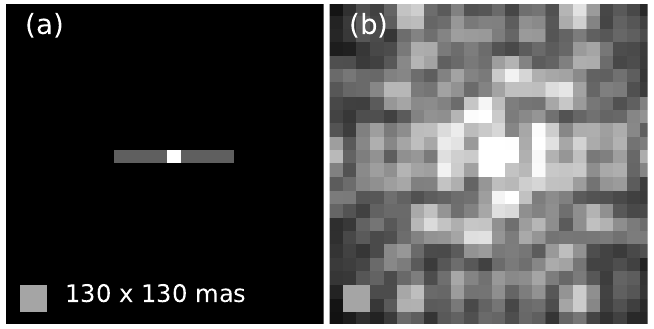}}
\caption{
Fiducial toy model of extended emission and simulated fringes. a) A model of
extended emission around a point source,
shown at NIRISS detector resolution.  We have chosen a horizontal bar (one model for AGN fuelling), 
surrounding a 7.5~mag unresolved point source and ignore host galaxy background. The bar is 9 NIRISS pixels long
(585~mas), with integrated flux of 8.5~mag.  Note this image is shown for
reference; when simulating observations, we convolve our toy model with 
the 11x oversampled PSF, before adding the noisy point source and rebinning
fringes to the scale of the NIRISS detector (see Fig.~\ref{fig:psfcompare}).  
Thus, in simulated observations, the bar 
is 1/11th of a NIRISS detector pixel in width (but length as shown).
b) Simulated NIRISS AMI mode observation of the bar described in a).  Integration
time is 0.66 (0.42)~sec in F380M, F430M (F480M) filters, 
and noise sources are described in the text. Compare to point source
in Fig.~\ref{fig:psfcompare}c.
Source structure is not apparent by eye.
\label{fig:tgt}}
\end{figure}

Observations were simulated
on a grid 11x finer than NIRISS pixels.  To simulate a PSF calibrator observation, the 
noisy point source exposures were binned up
by a factor of 11, as shown in the last panel of Fig.~\ref{fig:psfcompare}.
We generated a model of extended structure around the point source AGN on the same 
oversampled grid; convolved the model image with a noiseless oversampled PSF. 
We then shifted the oversampled, convolved model image to match the pointing error of the
$N_{exp}$th exposure of the simulated noisy point source sequence. We then binned up the noiseless, 
shifted, convolved extended structure by a factor of 11 and added it to the noisy point source observation sequence 
(with matching pointing error). Our approach is simpler than simulating the point source together with
extended structure through the whole optical path including noise, since noise sources associated with the point source will
dominate.  Fig.~\ref{fig:tgt} shows our fiducial extended structure model 
and the resulting NIRISS AMI mode interference fringes. Our model is a horizontal bar, 9 NIRISS pixels long
(585~mas), with integrated flux of 8.5~mag, around a 7.5~mag point source AGN. 
We also considered models non-aligned with our detector grid, which showed similar behavior.

We used the first 28 noisy exposures of the point source as our PSF calibrator, so 
exposure times were 18.48 (11.76)~sec at one orientation 
for the F380M, F430M and (F480M) filters.  The remaining 9 sequences of 28 exposures
(with added extended structure and different noise realizations) were used to simulate science target exposures, 
so science exposure times were 166.32 (105.84)~sec at
one orientation for the F380M, F430M and (F480M) filters.

\subsection{Fourier plane operations}
\label{sec:fplane}
\begin{figure}
\centerline{
\includegraphics[angle=0,width=8.0cm]{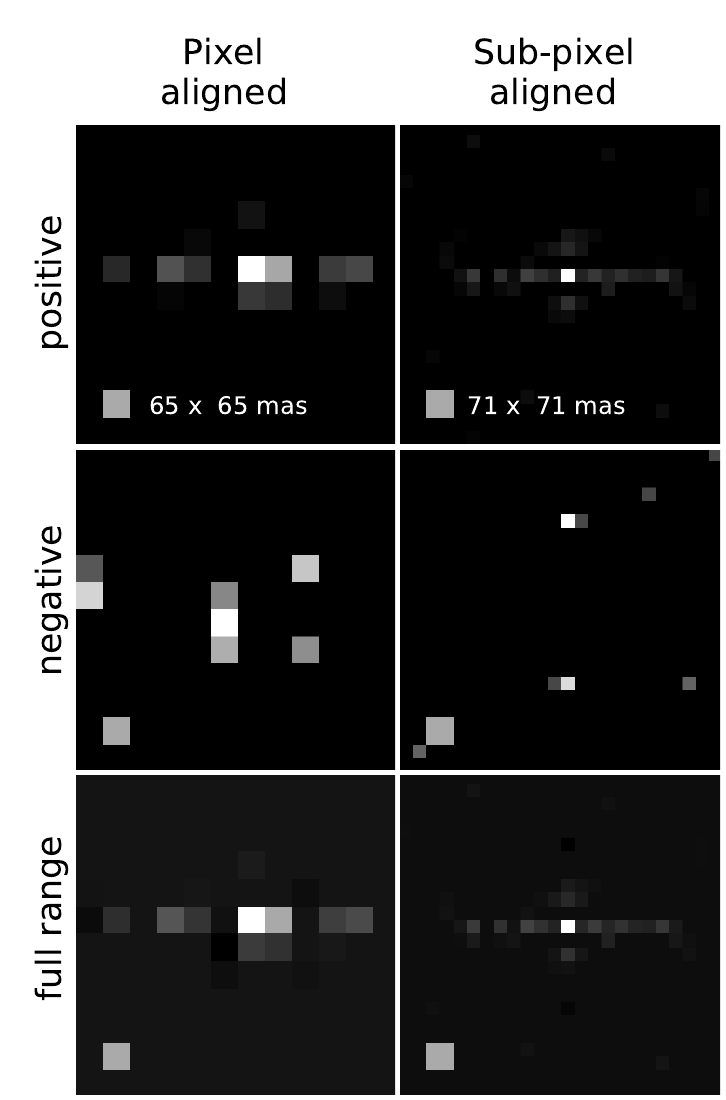}}
\caption{
\textsc{Clean} models for naive versus sub-pixel aligned deconvolution. 
All panels are displayed with a square root stretch.
Left panels: A simple \textsc{clean} deconvolution of
science target fringes (like those shown in
Fig.~\ref{fig:tgt}b), using IDL's \textsc{clean}.pro and point source calibrator
fringes (like those shown in Fig.~\ref{fig:psfcompare}c) for the
PSF.  Images were aligned to pixel accuracy.  Integration times are
2.8 (1.8)~min on target and 18 (12)~sec on the calibrator
in F380M, F430M (F480M) filters.  
At NIRISS pixel scale the bar should be only 9 pixels long (including the point
source).  Note also the relatively large negative artifacts near the point source. Though
pointing errors are small compared to a detector pixel, misalignments
between the target and calibrator are significant
enough to smear out the reconstruction.  Shorter integrations result in
less smearing but still produce large negative flux components and other artifacts
comparable to or brighter than the extended structure shown here.
Right panels: \textsc{Clean} model obtained using Fourier methods to
align images to sub-pixel accuracy, as well as the addition of an observation
at a second orientation (integration times are doubled from the
first panel) and using MIRIAD \textsc{clean}.  Pixels are $\sim$half the size of NIRISS, so the bar
should be 18 pixels long (including the point source). 
Gain (0.1) and number of iterations (250) in 
the \textsc{clean} loop was the same for
each panel. See text and Fig.~\ref{fig:hbarclmp} for further details
on the right panels.
\label{fig:naivecomp}}
\end{figure}

\begin{figure}
\centerline{
\includegraphics[angle=0,width=8.0cm]{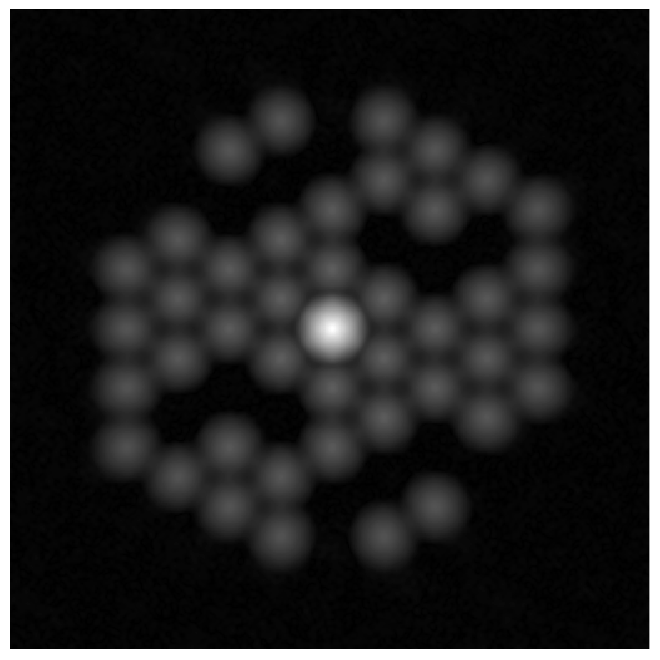}}
\caption{Amplitude splodges from a simulated 
noisy JWST-NIRISS AMI mode observation of a 7.5 magnitude point source 
with a 0.66 (0.41)~sec integration in F380M, F430M (F480M) filters, 
and noise sources as described in the text. These amplitudes come from the FFT of the image in  Fig.~\ref{fig:psfcompare}c, zero-padded by a factor of 4.
\label{fig:amp}}
\end{figure}

A simple \textsc{clean} \citep{hogbom} deconvolution (using IDL's \textsc{clean}.pro\footnote{http://www.boulder.swri.edu/$\sim$buie/idl/pro/clean.html}), is
shown in the left panels of Fig.~\ref{fig:naivecomp}.  Our pointing errors are large enough (at only 8\% of a NIRISS pixel) to introduce significant artifacts
in the reconstructed image. Instead we proceed by zero-padding the images and taking the FFT of all exposures (calibrator and science target).  We decompose 
the results into amplitude ($A(\bu)$, e.g. Fig.~\ref{fig:amp}) and phase ($\phi(\bu)$, e.g. Fig.~\ref{fig:pha}) of the complex visibility (for those new to interferometry,
see appendix \ref{sec:FT}). 
A change in telescope pointing introduces a uniform phase shift in the interference fringes (see appendix \ref{sec:tech}), while amplitudes are unaffected. In Fourier space this becomes a slope across the Fourier phase plane. Therefore if our exposures are short enough to be characterized by a single pointing, 
we can fit and subtract a plane from the raw Fourier phases ($\phi$) of each 
exposure; this yields sub-pixel aligned phases ($\phi^{\prime}$). For convenience, we co-add visibilities for the calibrator and science
target exposures respectively, forming a single set of sub-pixel aligned visibilities for each observation.

\begin{figure}
\centerline{
\includegraphics[angle=0,width=8.0cm]{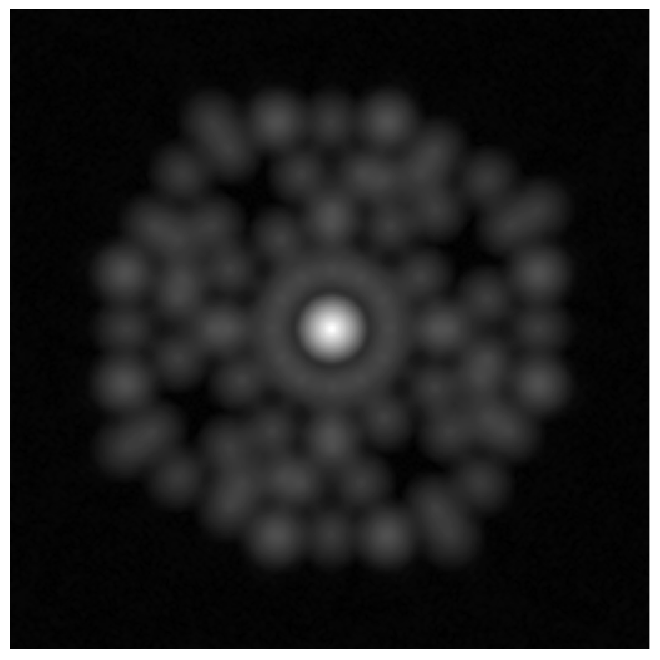}}
\caption{
Greatest $\bu$ plane coverage possible given 2 observations
separated by a roll angle of 90 degrees. Two sets of amplitude splodges
(as for figure~\ref{fig:amp}) with the second rotated by 90 degrees to the
first. Extraction reduces the coverage further (see figure \ref{fig:uvextract}, last
panel, for comparison, and appendix \ref{sec:tech}).
\label{fig:2orientcover}}
\end{figure}

Figs.~\ref{fig:amp} and \ref{fig:pha} show our maximum
$\bu$ plane (Fourier domain of our images) coverage at one orientation; note the gaps in coverage. Our $\bu$ plane is equivalent to the radio interferometric $(u,v)$ plane. Fig.~\ref{fig:2orientcover} shows the improved coverage for combined observations at two orientations, at $0^{\circ}$ and $90^{\circ}$, corresponding to JWST observations separated by a few months. We simulated observations at the second orientation by generating an identical extended source image rotated by the roll angle and proceeded as above. We then rotate the visibilities in the $\bu$ plane by the roll angle. For the rotationally symmetric point source calibrator, we simply rotated the visibilities. Note that the roll angle (or, equivalently, the timing of the second visit) does not require fine-tuning. Of course, for future space missions, if two complementary masks were available in a pupil wheel, near-simultaneous, complete $\bu$ plane coverage could be achieved. 

\begin{figure*}
\centerline{
\includegraphics[angle=0,width=18.0cm]{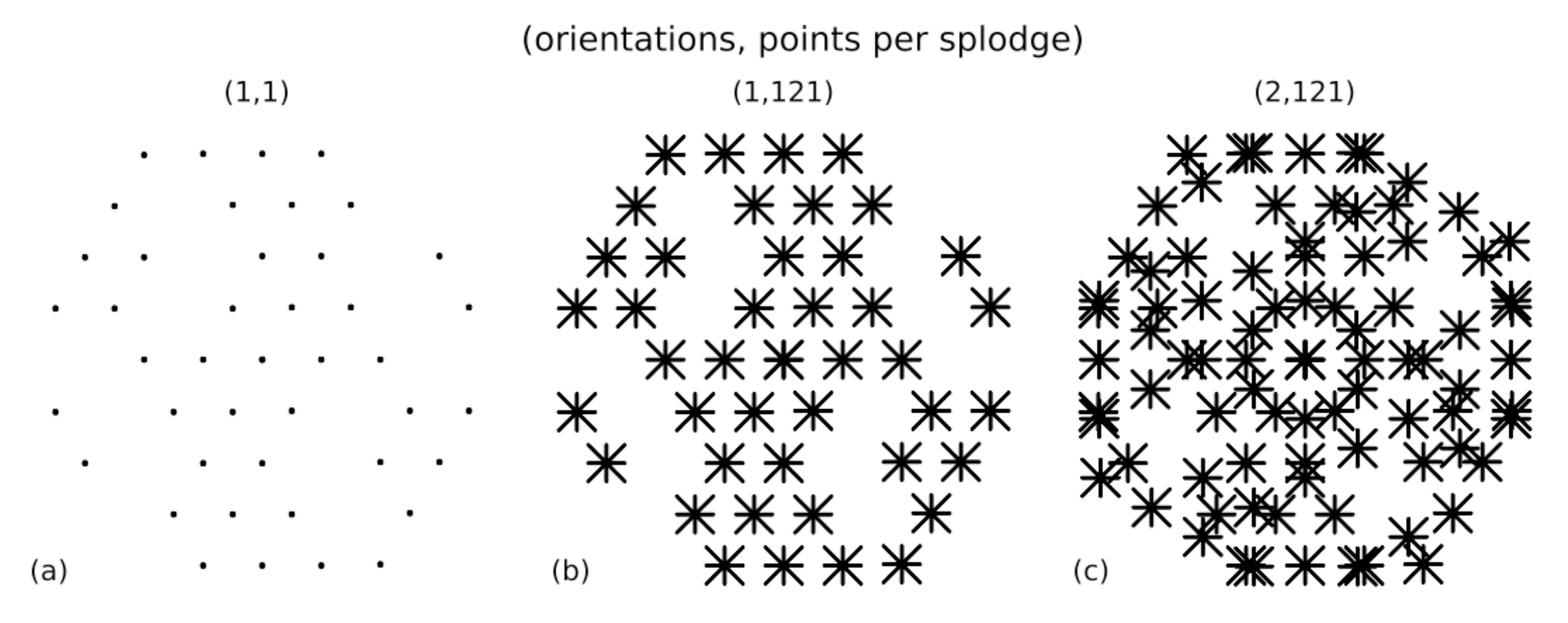}}
\caption{
Fourier ($\bu$) plane coverage after extraction. a) Extraction at one orientation,
using only $\bu$ splodge centers. b) Extraction at one orientation,
but using 15x8 additional points per splodge, 
arranged as the vertices and bisectors
of squares centered on the splodge; each square is 
two pixels larger than the previous one. c) Our standard
extraction, as for panel b) but observing at 2 orientations.
See appendix \ref{sec:tech} for further details.
\label{fig:uvextract}}
\end{figure*}

Note the pattern of extended `splodges' in the $\bu$ plane, 
rather than points or tracks (as in radio interferometry). The total number of independent data points in Fourier space is equal to the number of pixels
in image space (without windowing). Visibilities at splodge centers (e.g. Fig.~\ref{fig:uvextract}a) provide the 
only fully independent (non-redundant) information, but away from the centers 
there is additional, partially independent information. This additional information 
is not usually extracted in ground based O/IR observations (see appendix \ref{sec:tech}). 
By extracting the visibilities at points away from the splodge centers 
(i.e. away from those corresponding to the aperture center-to-center baselines) 
we get more imaging information in an observation than is possible from the 
ground. Target and calibrator visibilities are extracted at identical $\bu$ 
coordinates, chosen by analyzing the calibrator data. Fig.~\ref{fig:uvextract}c 
shows the $\bu$ extraction pattern used to generate the bottom-right panel of 
Fig.~\ref{fig:naivecomp}.  Appendix \ref{sec:tech} discusses the effect 
of different extraction patterns on image reconstructions.

\begin{figure}
\centerline{
\includegraphics[angle=0,width=8.0cm]{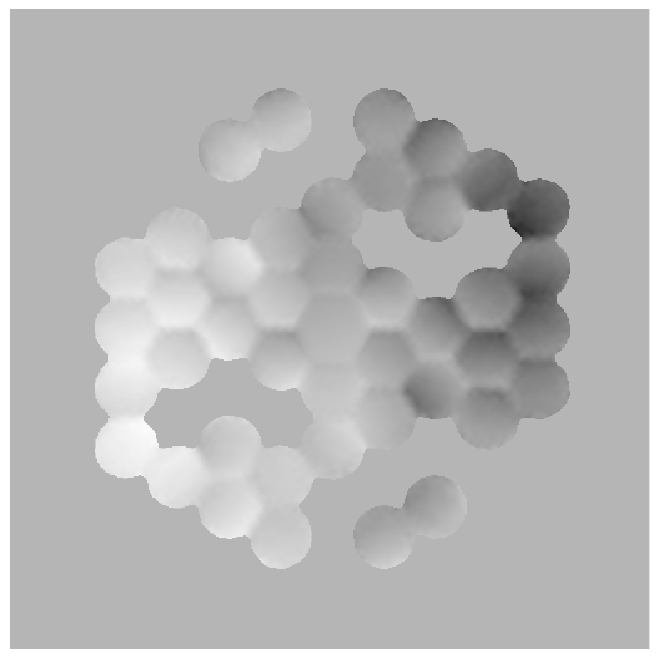}}
\caption{As Fig.~\ref{fig:amp}, except for phase splodges. Note that
in ground-based O/IR, fringe phases cannot be obtained, so closure phase and source priors must be used to obtain astrophysically interesting information.
\label{fig:pha}}
\end{figure}

\subsection{Deconvolution}
We pass the sub-pixel aligned, extracted visibilities to MIRIAD \citep{MIRIAD}, 
a radio interferometric software package.  We concatenate visibilities 
from multiple orientations, and separately `invert'
(inverse Fourier transform) calibrator and target data, as shown in  Fig.~\ref{fig:dirtymaps} (see
appendix \ref{sec:tech} for details).
We have recovered our original fringes, now properly
aligned for deconvolution, and weighted according to their
$\bu$ plane coverage (the weighting is non-uniform due to the combination of multiple
orientations and the lower weighting of off-center extracted points).  We have 
chosen an image plane pixel scale approximately twice as fine
as our original detector images (35.6~mas).
Pixel scale and signal-to-noise ratio (SNR) per pixel can be traded in any deconvolution scheme--this
is one effect of various weighting schemes \citep[see][and appendix \ref{sec:tech}]{Briggs}.
While this trade also exists for full-aperture imaging, the inherently smaller
PSF core and consequent improvement in theoretical angular resolution of an NRM 
system means that any result achievable
with a full-aperture deconvolution can be improved upon in NRM imaging.

We use the normalized inverse Fourier transform of the extracted, sub-pixel aligned
calibrator visibilities
as our PSF (or `dirty beam', $B$).  This contrasts with the standard radio strategy of
using the inverse Fourier transform of the target visibilities, with $(A,\phi)=(1,0)$ as the `dirty beam' or PSF.
Because we include $\bu$ points away from splodge centers, 
our technique is equivalent to constructing a full Spatial Transfer Function.
We use our PSF to \textsc{clean} the inverse Fourier transformed, extracted, sub-pixel aligned target visibilities (or `dirty map', $I^{\prime}_{\rm fringe}$).
We choose the smallest useful \textsc{clean} box. For the cases we show this was a square region 32 pixels on a side,
centered on the brightest pixel in the fringe image.  If we wish to image
larger regions, we can do so up to a region one-quarter the size of the
PSF; alternatively, we can use multiple pointings to construct
a mosaic of extremely large fields.

We show the \textsc{clean} model and restored map in 
Fig.~\ref{fig:hbarclmp} (see appendix \ref{sec:tech} for definitions).  The \textsc{clean} model is the same as that shown
in the bottom right panel of Fig.~\ref{fig:naivecomp} and compares well with the input model
shown in Fig.~\ref{fig:tgt}a.
For NRM systems which capture fringes as gridded detector images (like NIRISS), 
the most useful interpretation of the data is
obtained by viewing and comparing \emph{both} the \textsc{clean} model and the restored map. 

\begin{figure}
\centerline{
\includegraphics[angle=0,width=8.0cm]{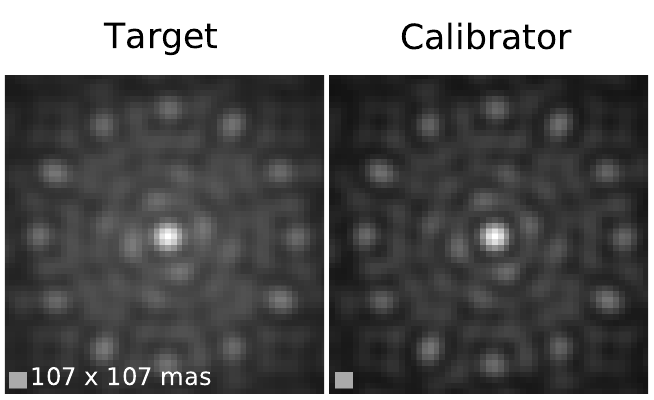}}
\caption{
Science target and calibrator fringes shown on square root stretch. 
a) Left panel shows inverse FT of sub-pixel aligned, extracted visibilities of observations at 2 orientations of our fiducial bar model.
The target was observed for a total of 5.5 (3.5)~min (half at each orientation) with 
NIRISS centered on $3.80,4.30,(4.80)\micron$ respectively. Pixel scale is twice as fine
as the JWST-NIRISS detector pixel scale.  b) Right panel shows the same
but for the point source calibrator observations.  Total exposure time is 37 (24)~sec with NIRISS, centered 
on $3.80,4.30,(4.80)\micron$ respectively.  This is our PSF.
Note the structural similarities between the two panels, but the left panel shows 
significant `filling in' of flux in the darker regions, due to the extended structure.
\label{fig:dirtymaps}}
\end{figure}

\begin{figure}
\centerline{
\includegraphics[angle=0,width=8.0cm]{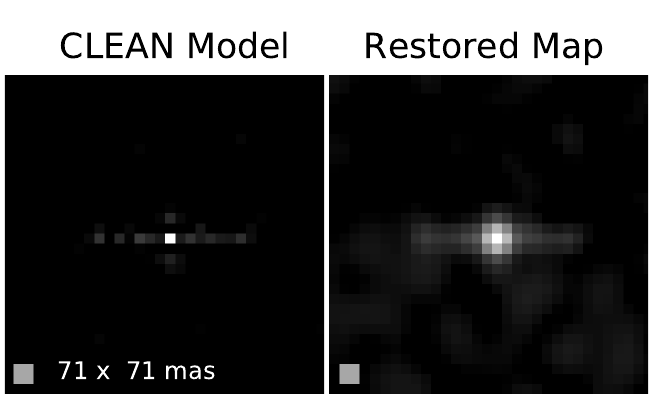}}
\caption{
\textsc{Clean} model and restored map from simulated observation of target shown in Fig.~\ref{fig:tgt}a.
a) Left panel shows MIRIAD \textsc{clean} output of Fig.~\ref{fig:dirtymaps}a,
using Fig.~\ref{fig:dirtymaps}b as the PSF.  Note that when we inverse FT 
our visibilities, we chose a pixel scale twice as fine as the JWST NIRISS detector pixel scale, so while the input bar (in Fig.~\ref{fig:tgt}) is 9 pixels long, a perfectly recovered \textsc{clean} model should be 18 pixels
long.  Our reconstructed bar is 17 pixels long (including point source), and recovers  $>90\%$ of the extended structure flux.
b) Right panel shows the restored map from MIRIAD \textsc{restor}, using a
symmetric Gaussian of FWHM=71~mas, for the restoring beam.  See
appendix \ref{sec:tech} for further details of \textsc{clean} and \textsc{restor}.
\label{fig:hbarclmp}}
\end{figure}

\begin{figure}
\centerline{
\includegraphics[width=9cm]{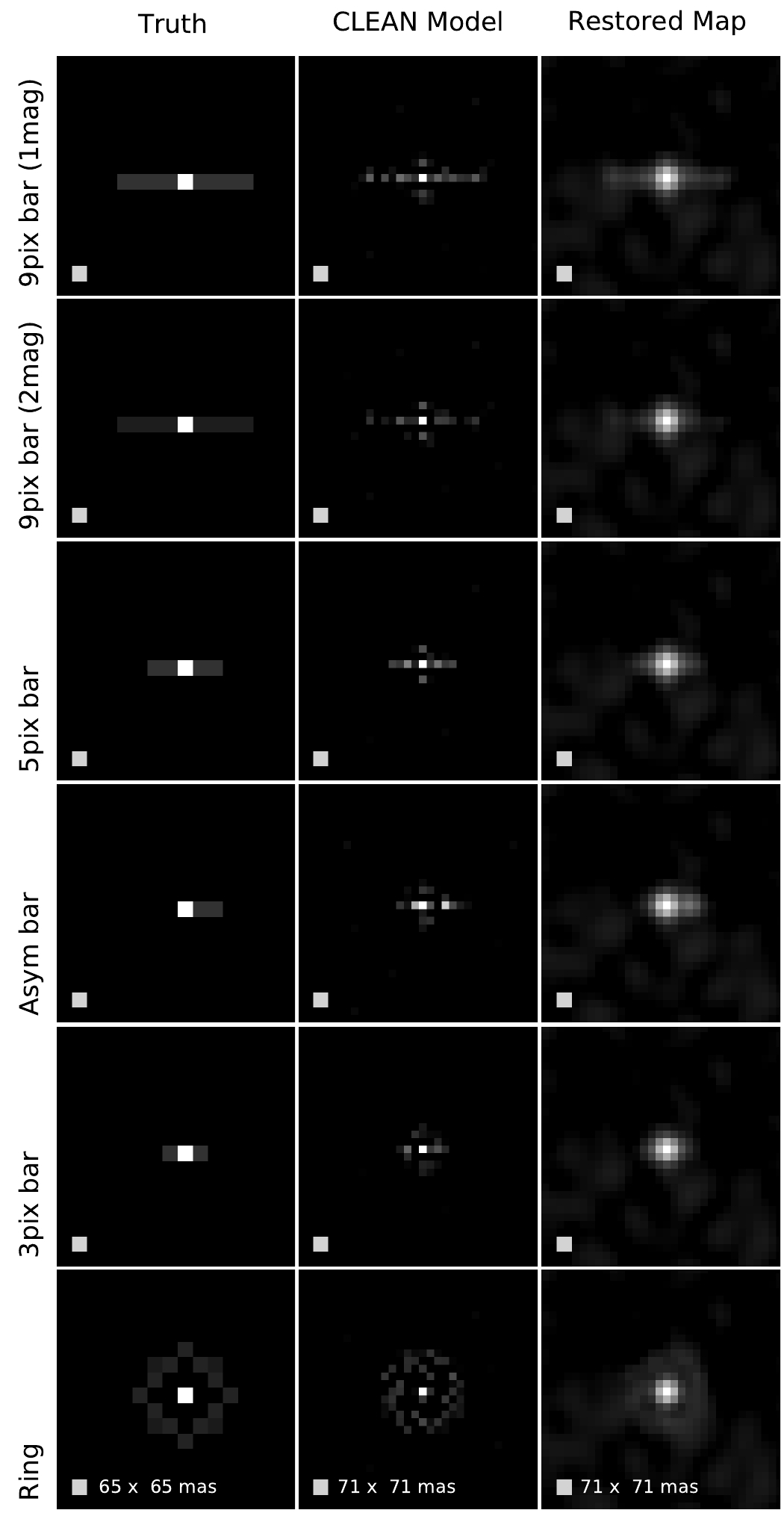}}
\caption{Input model, \textsc{clean} model and restored map for a variety of sky geometries.
All images displayed on a square root stretch.
First column is input model at NIRISS resolution (65~mas per pixel), 
second column is resulting \textsc{clean} model and third column is the restored map, both with pixels
half the size of NIRISS detector pixels.  Features in the first column should be same linear size in columns 2 and 3.
First row: same as Fig.~\ref{fig:tgt}a and \ref{fig:hbarclmp}.
Second row: as for first row but the integrated flux from the bar
is 9.5~mag (2~mag fainter than the point source).
Third row: horizontal bar 5 NIRISS pixels long (325~mas), with the same surface
brightness as the bar in the first row.
Fourth row: similar to the third row, but with the left side of the 
bar missing.
Fifth row: horizontal bar 3 NIRISS pixels long (195~mas), with the same surface
brightness as the bar in the first row.
Sixth row: circular ring of diameter 5 NIRISS pixels (325~mas).  Ring has
an integrated flux of 8.5~mag.
\label{fig:multmods}}
\end{figure}

\emph{Other models:} Fig.~\ref{fig:multmods} shows image reconstructions for some other extended 
emission models, including short and asymmetric bars and a ring (using the same
techniques and exposure times as above). The second row
demonstrates the recovery of a 9 NIRISS pixel long bar, at an integrated flux of 9.5~mag, or
2 mag integrated contrast, equivalent to 4.4 mag pixel-to-pixel contrast ($\sim10^{-2}$), in
5.5 (3.5)~min at $3.80,4.30,(4.80)\micron$. Assuming SNR improves as $\sqrt{t}$
(i.e. we are photon noise limited), we should reach 5~mag of 
contrast (pixel-to-pixel) imaging extended
structures in less than 10 minutes.

All results presented thus far use \textsc{clean}, without the use of a model prior
or \textsc{selfcal} (which would incorporate closure phase and closure amplitude 
information--see appendix \ref{sec:cpca} for definitions).
Much of ground-based optical interferometry is used to measure a small number
of parameters e.g. separation, contrast and position angle for two
point sources. However, this approach 
will not work for sources with complicated extended structure, such as AGNs, 
where prior models are extremely diverse and have large numbers of variables.
We note however that more advanced image reconstruction techniques are available 
\citep[][and references therein]{bar12}. 
We have not investigated their use on space-based data and offer our 
simulations as a conservative estimate of extended source
imaging capabilities.

\subsection{Other strategies}
\begin{figure*}
\centerline{
\includegraphics[width=18cm]{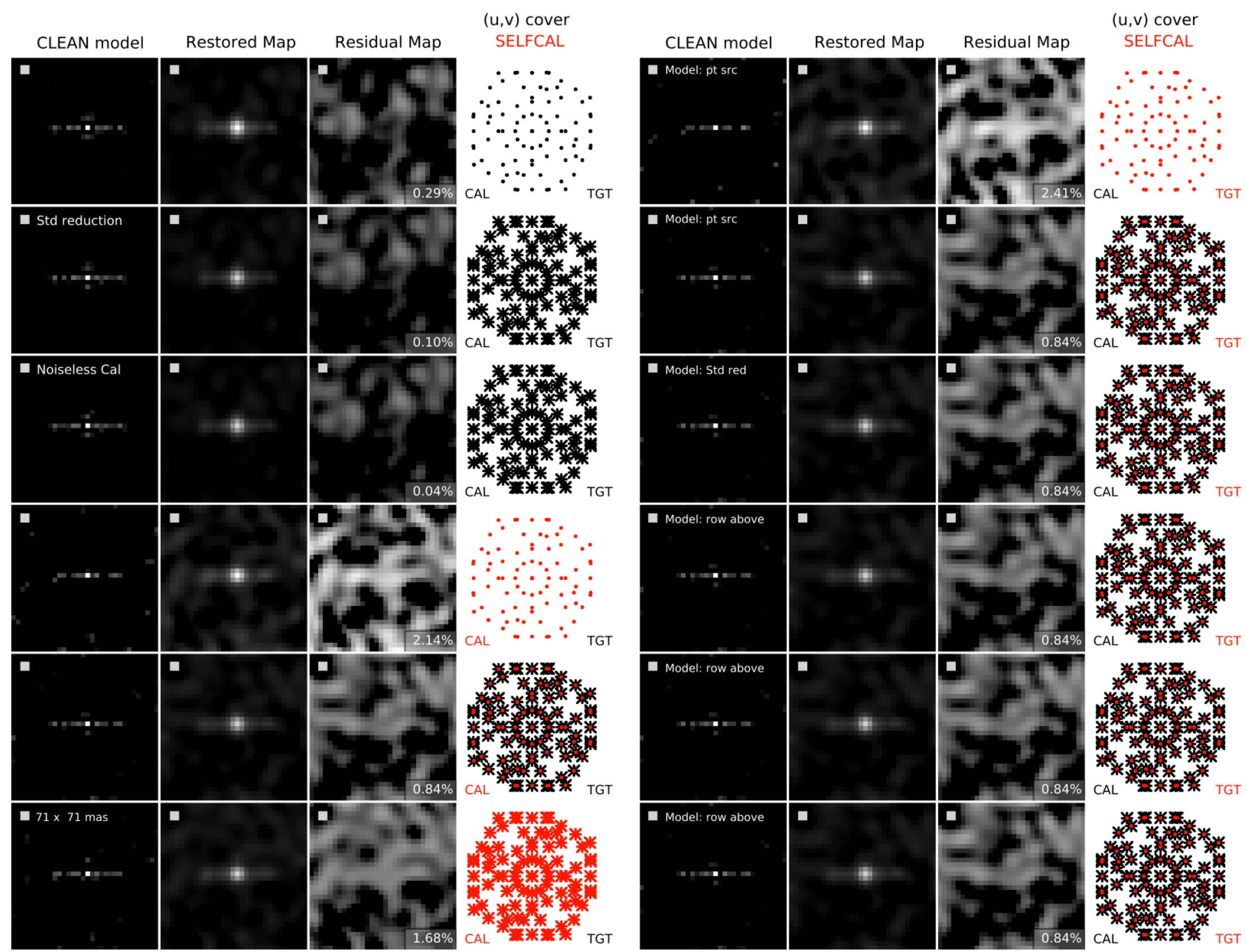}}
\caption{\textsc{Selfcal} experiments, with non-\textsc{selfcal} comparisons.  All experiments used our fiducial
bar for a science target. Fourier plane coverage is identical between calibrator
and science target observations for a given experiment; red points in the (u,v) cover column indicate \textsc{selfcal} was applied to those points for either 
the point source calibrator (red CAL) or the science target (red TGT). The percentage of flux remaining in 
residuals relative to the flux in the original fringe image is displayed in the bottom right corner of the
residual map.  All images are shown on a square root stretch; the residual maps are shown with the same range, so fainter features
from frame to frame indicate smaller residuals. Many \textsc{selfcal} residuals show significant bar structure.
\emph{Left columns, first 3 rows (no \textsc{selfcal}):} Increasing $\bu$ coverage improves our image 
reconstruction, reducing artifacts, residuals and more fully recovering the
bar. Using a noiseless calibrator slightly improves residuals.
\emph{Left columns, last 3 rows (\textsc{selfcal} on PSF):} Using \textsc{selfcal} decreases the number and brightness of artifacts immediately
adjacent to the point source in the \textsc{clean} models, but increases the number and brightness of scattered artifacts.  Residuals are also higher.
\emph{Right columns (\textsc{selfcal} on science target):} \textsc{Selfcal} input model is indicated for
each row at the top of the \textsc{clean} model (bottom 4 rows are an iterative sequence).  
Artifacts and residuals behave similarly to \textsc{selfcal} on PSF reductions.
\label{fig:selfcalexpts}}
\end{figure*}

We have investigated some extensions to the simplest \textsc{clean} reductions, shown
in Fig.~\ref{fig:selfcalexpts}.
\textsc{Selfcal} enforces closure relations on visibility data when a model prior allows
calculation of `expected' closure phases (CPs) and closure amplitudes (CAs). It assumes all departures from expected
relations are due to uncorrelated, aperture-specific errors, and computes a complex `gain' for each aperture
which corrects for those errors.
For our science
targets, we cannot guarantee the existence of such priors; however, our calibrator
is known to be a point source, so applying \textsc{selfcal} to that data might be expected to improve
our PSF measurement and the SNR in our image reconstructions.  Unfortunately, CP and CA relations
are not correctly defined by MIRIAD's \textsc{selfcal} for $(u,v)$ points away from splodge centers (see appendix \ref{sec:cpca}
for details). In addition, errors in our simulated data are not aperture specific, so unlike in the radio,
it is not sensible to apply the `gains' derived from our calibrator observations to our 
science observations (this may change in actual operations).
Fig.~\ref{fig:selfcalexpts} shows the \textsc{clean} model, restored map, residual map and $\bu$ plane coverage for a series
of experiments, all using our fiducial input model of a horizontal bar 585~mas long, with an integrated
flux of 8.5~mag, surrounding a 7.5~mag point source.  \textsc{Selfcal} experiments are indicated
in red (CAL if used on the point source calibrator or TGT on the science target), while comparison results 
without \textsc{selfcal} are indicated in black. 
We also list the (absolute value of) residual flux as a percentage of the 
(absolute value of) flux in the input fringe image.
We include a reduction using a simulated noiseless 
point source for a PSF--note the residuals decrease only slightly from our standard reduction. 

As can be seen in Fig.~\ref{fig:selfcalexpts}, application of \textsc{selfcal} does not improve
our image reconstructions, and in many ways makes them worse.  
Importantly, our source structure is plainly visible in many of the residual maps where 
\textsc{selfcal} was applied.  In general, applying \textsc{selfcal} to
our PSF improves our characterization of the PSF core (or main beam of our dirty beam).  This reduces
or eliminates the artifacts in our \textsc{clean} models appearing nearest the point source.  However,
\textsc{selfcal} significantly increases the noise level in the wings of the PSF (sidelobes), as seen
in the increased level of artifacts distributed throughout the \textsc{clean} models and
the increase in residual flux; the bar
does not stand out as strongly in the restored maps. Although
it is mathematically incorrect to do so, we did try to apply \textsc{selfcal} to splodge center \emph{and} offset visibilities with no improvement.
When applying \textsc{selfcal} to our science target, we must decide on a prior--we tried 1) a point source,
2) the output of our `blind' standard reduction, and 3) an iterative approach. 
When modelling the science target as a point source, \textsc{clean} still recovers a bar
feature, albeit with higher residuals than our standard reduction. In real operations, such an exercise might give additional 
confidence that features in a recovered image were not artifacts.
The iterative sequence uses the output of our standard reduction (appropriately clipped) as our input model for \textsc{selfcal}.
The following row uses the clean model from the row above (again clipped) as the input model for \textsc{selfcal}, and so on down the columns.
While we might hope such an approach would allow us to
improve our science image even in the absence of priors, Fig.~\ref{fig:selfcalexpts} implies that
this will not be possible for sources requiring significant dynamic range.  Artifacts quickly appear at
levels close to that of the extended structure (probably this results from the same core/wings noise shift seen in our
calibrator experiments); the clip level required to exclude artifacts also excludes
large portions of the real structure, making improvements over multiple iterations negligible.

\begin{figure}
\centerline{
\includegraphics[width=9cm]{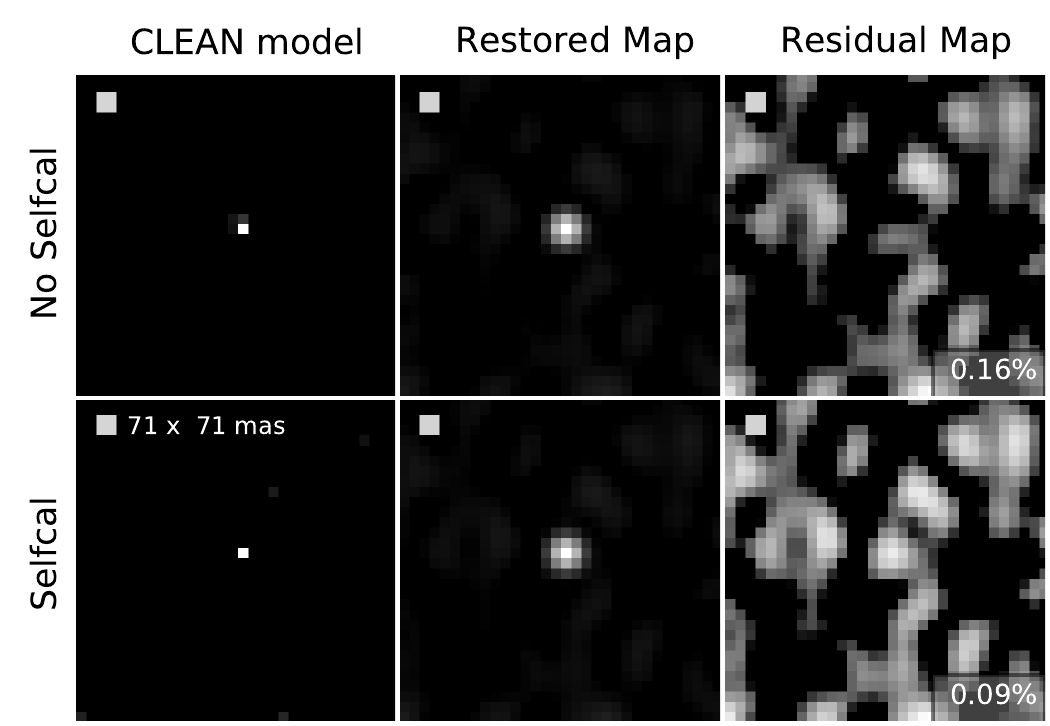}}
\caption{Noiseless point source science target. We show the \textsc{clean} models, restored maps and residual maps for simulated
observations of a noiseless point source, using our standard noisy point source observation as the PSF ((u,v) coverage is as for our standard
reductions).  We show results
both with and without the use of \textsc{selfcal}.  The level of residuals is comparable to that found
in our standard reduction, and implies that our noise floor may be set by noise in our PSF.  On orbit, this could be mitigated by longer
calibrator observations and/or use of noiseless simulated PSFs.  As in Fig.~\ref{fig:selfcalexpts}, use of \textsc{selfcal}
increases the level of scattered artifacts in the \textsc{clean} models, but decreases them immediately surrounding the point source.
\label{fig:noiseless}}
\end{figure}

We also simulated a noiseless observation of a point source, and used our standard (noisy) calibrator
observations to \textsc{clean} them.  The results are shown in Fig.~\ref{fig:noiseless}; we note that the level
of noise in the restored map is about $10^{-2}$ of the point source, and suspect that noise in our calibrator
observations is presently limiting the contrast achievable via imaging.  If JWST 
is sufficiently stable and well-characterized on-orbit,
we may be able to acheive higher contrasts using a noiseless simulated PSF; even if calibrator observations
are required (as is likely), longer calibrator integration times should improve achievable contrasts (also see the smaller
residuals in the Fig.~\ref{fig:selfcalexpts} panels where a noiseless point source was used for the PSF).

\subsection{Comparison with Ground-based systems}
\label{sec:groundcomp}
Ground-based O/IR interferometers
cannot measure absolute phase without resorting to exceedingly short (millisecond) exposures for atmospheric freezing
\citep[e.g.][]{han87}. Because of read-noise, this limits ground-based phase measurements to
the very brightest objects. As \citet{MonnierAllen2012} state: `Without valid phase 
information accompanying the visibility amplitude measurements, one cannot carry out the inverse
Fourier Transform that lies at the core of synthesis imaging and the \textsc{clean} algorithm specifically.'

The `images' produced by ground-based O/IR interferometers depend 
on closure phase to partially recover otherwise inaccessible phase information. 
Instead of inverse Fourier Transforming the fully measured complex visibilities
(as we do with AMI), ground-based O/IR imagers rely on modeling the
amplitude and closure phase expected given a prior model of the source.
Importantly, meaningful results depend on the accuracy of the model,
where the number of free parameters is fewer than the number of equations
available. An interferometer with $N$
apertures generates $N \choose 2$ independent phases, but only $N-1 \choose 2$ independent closure phases \citep{readhead88}. As $N$ grows large, this discrepancy grows smaller, but for e.g. VLT's NaCo SAM with $N=7$, closure phase only recovers 71\% of the phase information. 
For AGNs, where environments are likely to be complicated, and not well modeled by simple \emph{a priori} models with few parameters, this is a major barrier to advances from the ground.

Space holds other advantages, particularly for AGN science. Without phase, there can be no absolute astrometry,
only relative astrometry. If we wish to compare e.g. the geometry of a torus or mass reservoir in the IR with the
geometry of a jet from the radio (to look for differences in alignment), 
absolute astrometry is required (radio interferometers
do get phase and absolute astrometry from the ground).  Looking for offset
AGNs also requires absolute astrometry.

The phase problem arises again if we wish to perform very wide-field mosaicking.
Mosaic fields are limited to the size of the isoplanatic patch
using ground-based optical interferometers (due to atmospheric instability and the guide star
field-of-view limitations of AO systems \citep{Fried}).  At L-band this is slightly larger
than half an arcminute \citep{MonnierAllen2012}.
Very wide-field mosaicking is more likely to be used for non-AGN science
(though mapping of large scale jet interactions might call for such a technique), but we note
the issue here for its importance to other subfields (e.g. ISM). The size of the patch also
decreases at shorter wavelengths so mosaicking could be a critical advantage for space-based NRM
observations at optical wavelengths in future missions.

There are also advantages to measuring amplitudes from space. Our amplitudes
are measured with a higher signal-to-noise ratio than for similar ground-based systems,
as expected given the greater stability of space-based systems and the lower
thermal background.  Specifically, we expect $\sim 10^{-14}$~thermal photons~s$^{-1}$~pixel$^{-1}$ at
F480M, based on an operating temperature of $50$~K and a circular mirror of radius $3.25$~m.
Also, space-based systems can, uniquely in the O/IR, measure closure amplitude. 

\begin{figure}
\centerline{
\includegraphics[width=9.0cm]{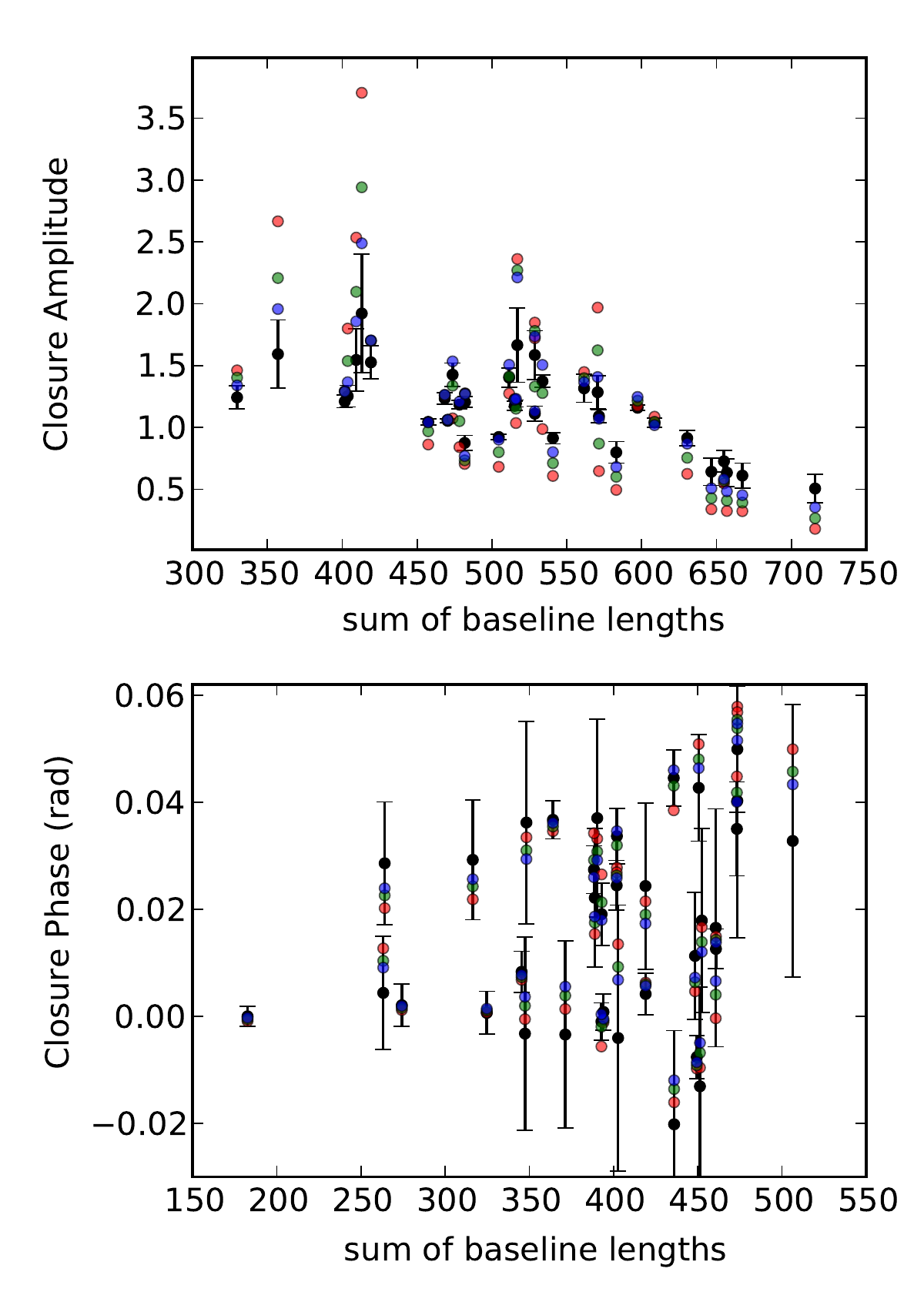}}
\caption{Closure phase (in radians) and closure amplitude for a 7.5~mag point source (black) and noisy observations of 585~mas horizontal bars 
surrounding a 7.5~mag point source at 1, 2, and 5~mag integrated flux contrasts (red, green and blue, respectively). Total baseline length is in arbitrary units.
Black circles represent CP (CA) values for a noiseless point source, which depart from theoretical expectations due to pixellation.  Error
bars are based on measured variance of CP (CA) for each baseline in 10 noisy point source observations, 
and show $1 \sigma$ departures corrected for the number of independent CPs and CAs.
\label{fig:cpca}}
\end{figure}

Though we do not use it in our imaging reductions,
we do measure closure amplitude (CA), as well as closure phase (CP).
These can be used to identify sources which depart significantly from point source
geometry; CA is particularly effective for this purpose.
We show in Fig.~\ref{fig:cpca} the closure phase and
closure amplitude, respectively, computed for our AMI observation
of a point source (black), as well as for extended source bar models 
like that shown in Fig.~\ref{fig:tgt} (1, 2, and 5~mag contrasts in red, green and blue, respectively--bars were observed for 2.8 (1.8)~min at F380M, F430M (F480M)). 
We plot CP (CA) against the total length of the baselines contributing to each aperture triangle (quad) in arbitrary units.
Though there are 35 triangles (quads) only 15 (14) are independent.
The black points are the values for a noiseless point source (which depart from theoretical expectations due to
pixellation); we computed CP and CA for 10 noisy realizations
of our standard point source calibrator observations (18.48 (11.76)~sec at F380M, F430M (F480M)) and 
found the standard deviation for each set of baselines.  The plotted error bars are
the usual standard deviation multiplied by 35/15 (35/14).

From the plot, we see that CP alone is barely able to distinguish the 1mag bar
from a point source; by contrast, CA alone is able to distinguish a 5~mag fainter (integrated flux 12.5~mag) bar
from a point source in less than 3 minutes.  Performing a chi-squared test on the data, we find the 5~mag contrast bar data
inconsistent with a point source at $5\sigma$ significance.
This corresponds to a point-to-point contrast of $\sim 10^{-3}$, equivalent to the
binary detection ability of VLT NaCo SAM \citep[e.g.][and note that binary
detection is an easier task than extended
structure detection because of the limited number of degrees of freedom in the system and
asymmetry, to which CP is more sensitive]{Lacour2011}. 
The remarkable stability of CA is what allows us to acheive this contrast.
Assuming the standard deviation increases with exposure time ($t$) only as $\sqrt{t}$
(i.e. if we are photon noise limited, which is likely to be true for the brightest targets) implies that a short $\sim$1.7 (1.1)~ks 
NIRISS AMI exposure, could distinguish an extended bar at 7 magnitudes total 
flux contrast from a point source at $5 \sigma$ statistical significance. 
This represents a point-to-point contrast of $\sim 10^{-4}$, or an order of magnitude 
better contrast than anything obtainable from ground-based O/IR interferometry.  We caution that
the chi-squared significance will be non-uniform depending on the science target structure and orientation.

\section{Example AGN targets}
\label{sec:sample}
\begin{figure*}
\centerline{
\includegraphics[angle=0,height=15.0cm]{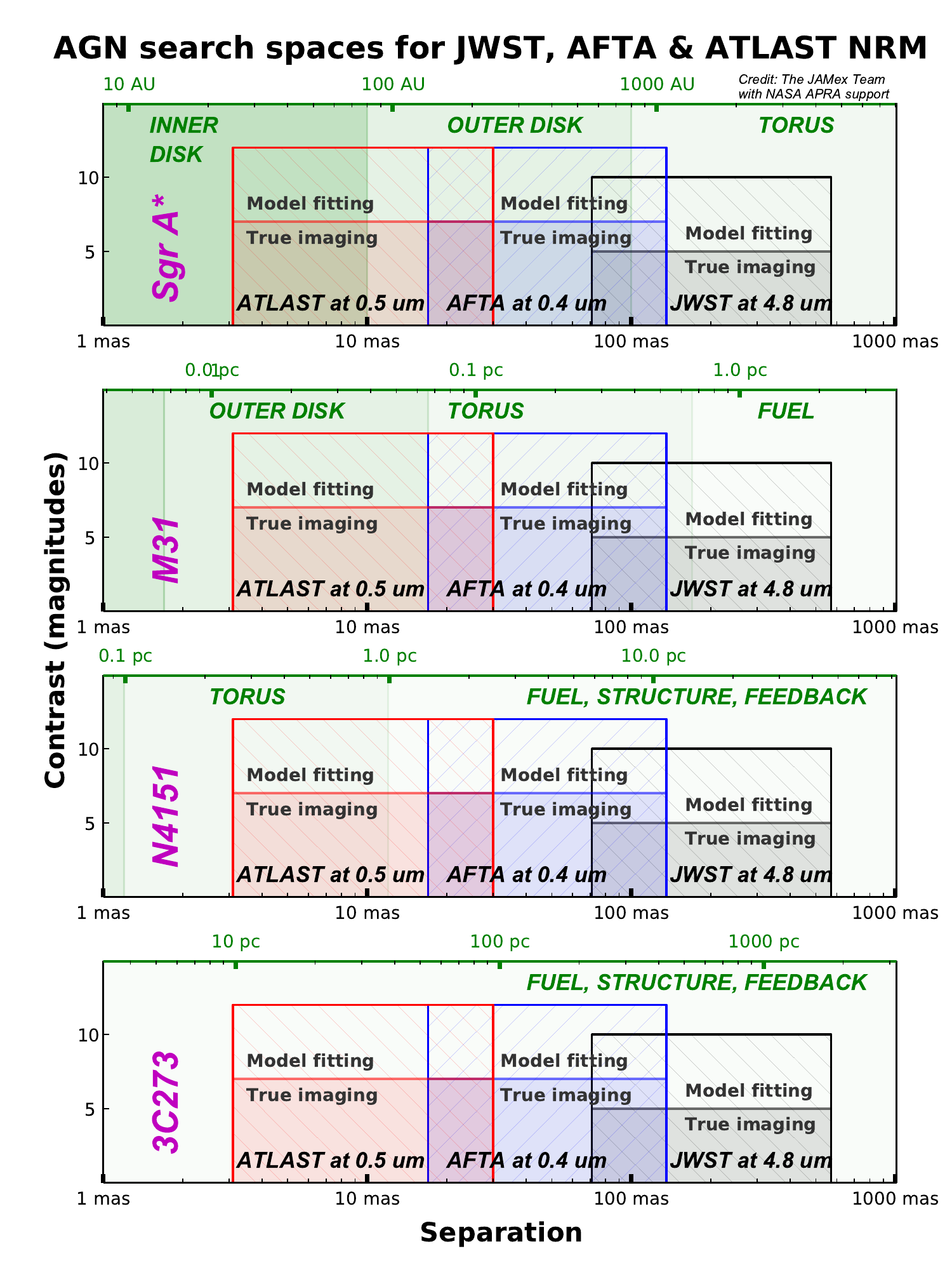}}
\caption{Distance scales (and regions of AGNs) that can be investigated using space-based 
aperture masking interferometry, for four fiducial accreting SMBHs from Table~\ref{tab:sample}. We
assume luminosity distances of 16.9~Mpc and 749~Mpc to NGC~4151
and 3C 273 respectively. Characteristic distance scales are distinguished and 
shaded as follows: Relativistic innermost accretion disk within $<100r_{g}$, Keplerian meta-stable
inner disk $10^{2}-10^{3}r_{g}$, outer disk $10^{3}-10^{4}r_{g}$, torus $10^{4}-10^{5}r_{g}$, region of 
AGN fuelling, feedback and binarity $>10^{5}r_{g}$. In each case we highlight the region of the 
galactic nucleus that could be probed by the three instruments in Table~\ref{tab:sample}. For JWST-NIRISS AMI,
the shaded region (`true imaging') corresponds to the imaging of extended structure at 5~mag pixel-to-pixel contrast ($10^{-2}$), surrounding a 7.5~mag point source, observed at 2 orientations for a total of 5.5 (3.5)~min using F380M, F430M (F480M) filters. The unshaded NIRISS region (`model fitting') corresponds to the $5\sigma$ detection of a modeled binary at 10~mag contrast near a 7.5~mag central point source \citep{siva09,siva10,siva12}. The shaded regions for AFTA (at 400nm) and ATLAST (at 500nm) correspond to assumed achieveable pixel-to-pixel contrasts of 
7~mag for extended structure, assuming photon noise dominates and extrapolating from JWST simulations. The unshaded 
AFTA and ATLAST regions correspond to the $5\sigma$ detection of a modeled binary at 
12~mag constrast near a 7.5~mag central point source.
\label{fig:multi}}
\end{figure*}

Table~\ref{tab:sample} lists targets for JWST-NIRISS AMI and future space-based NRM missions, 
spanning a range of activity types and black hole masses ($M_{\rm BH}$). Figure~\ref{fig:multi}
shows the science topics within reach of the various missions.
Since the size of an AGN central engine is supposed to scale
with $M_{\rm BH}$ \citep{ant93,urpad95,pet04}, NIRISS AMI
will probe different structures around nearby AGNs with $M_{\rm BH}$ 
spanning $10^{6-9}M_{\odot}$. Column 8 in Table~\ref{tab:sample} lists the regions which can be imaged with JWST-NIRISS AMI, spanning (IWA,OWA)=(75,600)~mas assuming $\lambda=4.8\mu$m. JWST-NIRISS AMI can image the outskirts of nearby AGNs as well as occasional flaring in otherwise quiescent nuclei (Sgr A*, M31). In nearby Seyferts, NIRISS AMI images will distinguish between models of continuous AGN 
fuelling (spiral or bar inflows from further out) and intermittent accretion events. Columns 9 \& 10 in Table~\ref{tab:sample} list region sizes that could be imaged with future missions: a UV/optical-band Astrophysics Focused Telescope Assets (AFTA) with a 2.4~m primary mirror ($\lambda \sim 400$nm) and Advanced Technology Large-Aperture Space Telescope (ATLAST), a proposed optical mission ($\lambda \sim 500$nm) with a 16~m primary mirror. Possible future missions are discussed further in \S\ref{sec:future}.
Here we discuss detailed science goals for NIRISS AMI in several archetypal sources: NGC 4151, 3C 273 and M31.

\begin{deluxetable*}{lrrcclrrrr}
\tablecaption{Galactic nuclei targets for NRM on Space Telescopes \label{tab:sample}}
\tablecolumns{10}
\tablewidth{0pt}
\tablehead{
\colhead{Name} & \colhead{RA} & \colhead{Dec.} & \colhead{$D_{L}$} & \colhead{log($M_{BH}$)} & \colhead{AGN} & \colhead{Apparent} & \colhead{JWST} & \colhead{AFTA} & \colhead{16m ATLAST} \\
& & & & & & \colhead{Magnitude} & \colhead{4.8$\mu$m} & \colhead{400nm} & \colhead{500nm} \\
& \colhead{(J2000)} & \colhead{(J2000)} & \colhead{(Mpc)} & & \colhead{class} & \colhead{(for aperture)} & \colhead{inner-outer} & \colhead{inner-outer} & \colhead{inner-outer} \\
 & & & &
 & & \colhead{(mag)($\arcsec$)} & \colhead{(pc)} & \colhead{(pc)} & \colhead{(pc)}  \\
 & & & & & & & & & \\
\hline
 & & & & & & & & & \\
\colhead{(1)} & \colhead{(2)} & \colhead{(3)} & \colhead{(4)} & \colhead{(5)} & \colhead{(6)} & \colhead{(7)} &\colhead{(8)} &\colhead{(9)} &\colhead{(10)} }\\
\startdata
NGC 4151 & 12 10 32.6 & 39 24 21 & \phn16.9\phn \phn & 7.1 & Sy 1& $6.6 \pm 1$\phm{0..0}(6) & 6.2--49.9\phn &1.41--11.3 & 0.3--2.11\\
NGC 4051 & 12 03 09.6 & 44 31 53 & \phn12.7\phn \phn & 5.1&NLSy1 & $7.9\pm 1$\phm{..0}(15) & 4.7--37.5\phn &1.06--8.5\phn & 0.2--1.63\\
NGC 3227 & 10 23 30.6 & 19 51 54 & \phn20.2\phn \phn & 7.6&Sy1.5 & $8.0\pm 1$\phm{..0}(15) &7.5--59.7\phn &1.7--13.5 &0.32--2.59\\
\\
M87          & 12 30 49.4 & 12 23 28 & \phn22.3\phn \phn & 9.8 &Lr,j& $7.5\pm 1$\phm{..0(00)} &8.2--65.9\phn & 1.9--14.9 & 0.36--2.85 \\
Cen A       & 13 25 27.6 & $-$43 01 09& \phn11.0\phn \phn &7.7 & Sy2 & $5.2\pm 0.2$\phm{.0}(5) &4.1--32.5\phn & 0.92--7.3\phn & 0.18--1.41\\
M81          & 09 55 33.2 & 69 03 55& \phn \phn0.7  \phn \phn &7.8 & L & $8.5\pm 1$\phm{.0}(3.5) &0.26--2.1\phn\phn & 0.06--0.5\phn &0.02--0.09\\
\\
Sgr A*       & 17 45 40.0 & $-$29 00 28& \phn \phn0.008 & 6.6 & \nodata & \nodata&609--4874AU & 138--1100AU & 26--205AU\\
M31          & 00 42 44.3& 41 16 09& \phn \phn0.8\phn \phn& 7.6 & L2 & $6.5\pm 0.2$\phm{.}(15) &0.30--2.36\phn & 0.07--0.53 & 0.013--0.10\\
\\
NGC 1068 & 02 42 40.7& $-$00 00 48 & \phn12.5\phn \phn & 7.2 &Sy 2 & $5.5\pm1.0$(0.3) & 4.6--37\phd\phn\phn &1.04--8.33 & 0.2--1.6\phn \\
NGC 4258 & 12 18 57.5 & 47 18 14 & \phn \phn9.0\phn \phn   &7.6 &Sy2,j & $\sim8.9$\phm{1.0.0}(7) &3.3--27\phd\phn\phn & 0.75--6.0\phn & 0.14--1.15\\
Circinus    & 14 13 09.9 & $-$65 20 21& \phn \phn8.0\phn \phn   &6.0 &Sy1h & $\sim4$--7\phm{1.0}(0.4) &3.0--24\phd\phn\phn & 0.7--5.3\phn & 0.13--1.02\\
NGC 4945 & 13 05 27.5 & $-$49 28 06& \phn11.1\phn \phn    &6.1 &Sy2 &$\sim8.3$\phm{1.0.}(14) &4.1--33\phd\phn\phn & 0.93--7.4\phn & 0.18--1.41 \\
\\
3c 273     & 12 29 06.7 & 02 03 09 & 749 \phn \phn \phn  &9.0 & QSO & $\sim7.8$\phm{1.0}(4.6) &277--2212\phd & 62--499\phd & 12--96\phn\phd \\
\enddata
\tablecomments{\scriptsize Potential galactic nuclei targets for NRM observations with JWST-NIRISS AMI and proposed future missions, spanning a range of AGN types, black hole masses, distances and broad sky
coverage. Column 1 is AGN name, column 4 is the luminosity distance (Mpc)
listed in NED. Column
5 is the logarithm of the black hole mass \citep{mck10a}.
Column 6 is the AGN classification(Sy=Seyfert, Lr=LINER, LL=low
luminosity AGN,j=prominent jet, Bl=Blazar, QSO=quasar). Sy1h denotes a
hidden Seyfert 1, visible
only in polarized light and NLSy1 is a narrow line Sy1. Column 7 lists the
range of observed L-band magnitudes of the AGNs (or as close as possible to  $4.5\mu$m), with (in brackets) the aperture size listed in arcseconds where available (from NED). Column 8 shows the distance scales (in pc except for Sgr A*), that can be imaged in the AGN and galactic nuclei for JWST-NIRISS AMI and correspond to inner and outer working angles of $\sim$75-600~mas. Column 9 shows the corresponding distance scales that could be imaged by a 2.4~m AFTA operating at $400$nm. Column 10 shows the distance scales that could be probed by a 16~m ATLAST mission operating at $500$nm (for an 8~m ATLAST, multiply distances by two).}
\end{deluxetable*}

\subsection{NGC 4151: archetypal Seyfert 1 AGN}
\label{sec:n4151}
NGC~4151 ($M_{\rm BH} \sim 1.3 \times 10^{7}M_{\odot}$) is the archetypal Seyfert 1 AGN\footnote{presently classified as a Sy1.5}. NGC~4151 is bright ($\sim 10^{44} \rm{erg} \rm{s}^{-1}$ bolometric luminosity), nearby (16.9~Mpc), hosted in a weakly barred spiral galaxy \citep{deV91}, and is the most studied AGN at most wavelengths \citep[e.g.][]{ulrich00}. The bar contains a 
total mass of $\sim 6 \times 10^{8}M_{\odot}$ in HI, much of which is inflowing
 \citep{ped93}. A radio jet spans $4\arcsec$ either side of the nucleus and is misaligned by $\sim 25^{o}$ with the bar \citep{hutch98,mund03}. How the misalignment occurs, and the structure and dynamics of gas within $\sim 10$~pc (150~mas) remains unknown.  Around $10^{9}M_{\odot}$ in gas lives within the central $\sim 900$~mas (60~pc) and $\sim 5 \times 10^{7}M_{\odot}$ within the central $\sim 150$~mas (10~pc) \citep{winge99}. From mid-IR interferometric observations there is structure on scales of $2.0 \pm 0.4$~pc ($\sim$30~mas) FWHM \citep{burt09}. However, this model-dependent estimate provides no information on geometry or clumpiness and insufficient size constraints. The structure could include part of the torus, or bars/spirals feeding the torus, or a warp from galactic plane to equatorial plane of the black hole \citep{bp75,kp06}. An AMI image (without priors) could distinguish between these scenarios and test models of feedback.
 
Assuming noise declines as $\sqrt{t}$, AMI imaging should probe
complicated extended structure without prior modeling at a
pixel-to-pixel contrast $\sim 10^{-3}$ in a  $\sim 33(21)$~ks total
exposure at two orientations, centered on
$3.80,4.30,(4.80)\micron$. At these contrasts, NIRISS AMI can
distinguish ordered fuelling structures from a disordered
environment. Previous studies provide us with `priors', which can
guide our interpretation of blindly deconvolved images. For example,
radio lobes at $\sim 200$~mas on either side of the NGC~4151 (right in
the NIRISS search region), may coincide with shocks where the jet
encounters interstellar medium \citep{mund03}. If a coincidental structure shows up in an AMI image \emph{in the correct location and orientation}, this radio prior enhances our ability to engage in physical interpretation.
Additionally, short exposure ($\sim 2$~ks) NIRISS AMI observations, can test for the presence of symmetric extended structure (bars, torus, rings, spirals), via closure amplitude, down to a pixel-to-pixel contrast ratio of $10^{-4}$  at $5\sigma$ statistical significance.

\subsection{3C 273: The original quasar}
\label{sec:3c}
3C 273 was the first quasar recognized to be extra-galactic \citep{sch63}, lying at $z=0.158$ (749~Mpc luminosity distance; $1\arcsec$=2.7~kpc linear scale) and is well studied at all wavelengths \citep[e.g.][]{sold08}. In most bands, the quasar core is saturated. So, modelling the underlying quasar environment and host galaxy requires PSF subtraction \citep[e.g.][]{bahcall95,hutch04}. \citet{bahcall97} find agreement with an underlying E4 host galaxy from fitting the residual intensity in regions $>1\arcsec$ after subtraction of a best-fit stellar PSF. However, any structure inside $1\arcsec$ will complicate both the subtraction of a stellar PSF and the model-dependent interpretation of the quasar host galaxy. The highest angular resolution study of 3C 273 to date was with the HST-ACS coronagraph \citep{martel03}, down to an IWA of 1300~mas. NIRISS AMI can probe a factor of 20 closer (75-600~mas) than the HST-ACS coronagraph (and a factor of 3 closer than 
coronagraphs on JWST). NIRISS AMI can test whether spiral structure 
at $\sim 1\arcsec$ which wraps around the quasar \citep{martel03} continues inwards, possibly fuelling the quasar. NIRISS AMI can also test a merger origin for the quasar in departures from a smooth light profile at $<1\arcsec$. Once again, previous studies can provide `priors' to guide deconvolution and physical interpretation. For example, \citet{homan01} measure the jet angle so structure that appears in a deconvolved image along the jet axis could be examined for a physical relationship to the jet. 

\subsection{M31: Nearby galactic nucleus}
\label{sec:m31}
M31 at $\sim 0.8$~Mpc is the next nearest SMBH ($1.4 \times 10^{8}M_{\odot}$) after Sgr A* but with less extinction along the sightline \citep{tempel10}. Within $3\arcsec$ ($\sim 10$~pc) of the center lies a double nuclear cluster of old red stars \citep{lauer98}.  The double cluster is modeled as a projection effect of an eccentric disk of stars about $\sim 2$~pc in radius ($\sim 600$~mas) \citep{trem95}, which may be non-aligned with the M31 disk plane \citep{peiris03}. Inside the double cluster at $\sim 1$~pc ($\sim 300$~mas) lies a Keplerian disk of blue stars $\sim 100-200$~Myrs old \citep{bender05,lauer12}. The origin of both disks is unknown. Observations with NIRISS AMI can constrain the inclination angle and surface brightness of the eccentric disk which will help us understand its dynamical evolution, its origin and possibly its interaction with the inner Keplerian disk. For future missions, an IFU combined with AMI on a space telescope would also extract velocity dispersion information for the stellar disk.

\section{Future missions: AFTA and ATLAST}
\label{sec:future}
NRM adds significant science impact to any proposed space telescope,
for minor technical considerations (i.e. requiring detectors to be
sampled at the resolution achievable with the mask, or pixellation of
at least $\lambda/4D$). Future missions with AMI mode capabilities
could probe regions deep in the potential well of SMBHs. For example,
an NRM on ATLAST could image close to the event horizon of Sgr A* and
could image to within a few hundred gravitational radii
($r_{g}=GM/c^{2}$) of nearby $\sim 10^{9}M_{\odot}$ Seyfert
AGNs. ATLAST AMI could therefore be used to image whole AGN disks in
the nearby Universe, allowing us to measure disk warps, gaps and
cavities directly. From Table~\ref{tab:sample} and
Fig.~\ref{fig:multi}, we see that a 16~m ATLAST is significantly
preferable to an 8m version, since it gives us most of the torus
region in nearby Seyferts (IWA $\sim 10$ vs $\sim 20$~pc) and better
access to the outer disk regions of some of the nearest SMBHs.

Figure~\ref{fig:multi} also shows the regions of several galactic
nuclei that could be probed with AFTA. A 2.4~m space-based telescope
used at optical and UV wavelengths ($\sim 400$nm) with an NRM, would
probe different structures than JWST's NIRISS AMI, particularly outer
disk structure in nearby massive Seyfert AGNs. It could also
investigate the connection between AGNs and star formation. For very
large aperture future missions (e.g. a 16~m optical ATLAST), the
additional collecting area would allow routine use of spectrometers in
AMI mode. Simultaneous spectral information would help constrain the
physical conditions and dynamics of imaged structures in
AGNs--important for answers to many of the big questions outlined in
\S\ref{sec:agn}. Sufficiently large collecting areas on future space
telescopes would also permit us to image structures in polarized light
from AGNs \citep{sivawp}.  Finally, while NIRISS AMI can expect to
observe tens of AGNs, future missions could target hundreds.

\section{Conclusions}
\label{sec:conclusions}
Space-based aperture masking interferometry (AMI) can image extended structures around AGNs and quasars at moderate contrast and high angular resolution. Space-based imaging does not require a prior model of the target (we obtain fringe phase) unlike ground-based O/IR interferometry. O/IR interferometry also cannot obtain closure amplitudes from the ground, due to atmospheric instability.  We demonstrate that a pixel-to-pixel contrast of  $10^{-4}$ should be achieved in short exposures with JWST-NIRISS AMI, an order of magnitude better than obtainable with long exposure ground-based O/IR interferometry. JWST-NIRISS AMI will be a unique
facility for carrying out moderate to high contrast observations of AGNs at 
high resolution, allowing us to constrain models of AGNs binarity, fuelling and structure at levels beyond present and proposed ground-based observatories. NRMs should be considered for addition to the filter wheel of most future space telescope missions, allowing high resolution and moderate to high contrast images of extended emission 
around bright sources in the optical/IR/UV bands. 

\section*{Acknowledgements}
The authors thank the referee for insightful and scholarly comments,
as well as helpful suggestions that improved the paper's clarity, completeness,
and organization.
This work is supported in part by the National Science Foundation grant
AST 08-04417, PHY 11-25915, the NASA grant APRA 08-0117, and the STScI Director’s Discretionary
Research Fund. KESF acknowledges the support of a BMCC Faculty Development
Grant.
We acknowledge many useful discussions with
Ron Allen, 
Julie Comerford,
Laura Ferrarese,
John Hutchings,
Jin Koda,
John Monnier, and
Peter Teuben.
KESF would like to dedicate this paper to her father, who passed away in June 2013,
after doing so much to help his daughter build a career in astrophysics.

\appendix

\section{Non-redundant masking overview} \label{sec:FT} 
We briefly review some relevant details of non-redundant masking (NRM) for
observers unfamiliar with the technique.  More complete descriptions of the
technique can be found in
\citet{mon03,MonnierAllen2012,ireland13}.  We assume
the Fraunhofer approximation of Fourier optics, \viz\ the image plane complex
amplitude $a(\bk)$ is the Fourier transform of the pupil plane complex
amplitude $A(\bx)$.  In the monochromatic case, at wavelength $\lambda$,
expressing pupil plane coordinates $\bx\equiv (x,y)$ in units of the wavelength
yields image plane coordinates $\bk \equiv (k_x,k_y)$ in radians.  We refer the
reader to \citet[\eg][]{bracewell} for details on Fourier theoretical results
utilized in this section.

While our simulations have focussed on JWST-NIRISS' 7-hole mask
(Fig.~\ref{fig:mask7fig}), we describe NRM in more general terms here.  $N_h$
identical holes in a pupil mask generate $N_h(N_h-1)/2$ sinusoidal fringes in
the complex amplitude of the image plane.  Each fringe is multiplied by an
envelope function, the \textit{primary beam},  which is the Fourier transform
of the individual hole transmission function.  The vector between a pair of
holes in the mask is referred to as a \textit{baseline}.  The image intensity
$I(\bk)$ for the JWST-NIRISS 7-hole mask is shown in
Fig.~\ref{fig:psfcompare}~(b and c).
$I(\bk)$ only contains spatial frequencies of the sky brightness distribution
that are transmitted by the pupil mask.  The \textit{complex visibility},
$\mathcal{V}(\bu)$, is the Fourier transform of the image intensity $I(\bk)$.
The domain of the complex visibility function is a pupil-like  `spatial
frequency' space, with coordinates $\bu \equiv (u,v)$.
$\mathcal{V}(\bu)$ is the Fourier component of the image brightness with
spatial frequency $\bu$.  An inspection of the absolute value of
$\mathcal{V}(\bu)$ shows that only certain spatial frequencies pass through the
mask (see Fig.~\ref{fig:amp}).  For a point source target, the quantity plotted
in this figure is commonly known as the \textit{modulation transfer function}
(or MTF) of the optical system.  Its support is obviously the region where
the autocorrelation function of the pupil mask is non-zero.  The isolated areas
of signal in this plane, \textit{splodges} \citep{lloyd06}, are
twice as wide as the hole size.  The centers of splodges are at the vector
separations of the hole centers, \ie\ the baselines.  Since $I(\bk)$ is real,
$\mathcal{V}(\bu)$ is Hermitian, so only one half of this space provides
independent information.  Thus, of the $7\times 6 = 42$ splodges in
Fig.~\ref{fig:amp}, only 21 are independent.  By the so-called Fourier `DC'
theorem, $\mathcal{V}(\bf{0})$ is the energy in $I(\bk)$ integrated over the
entire image plane.
Since no vector baselines in the pupil mask are repeated, each splodge is the
result of only one baseline.  It is this fact that enables unambiguous
calibration of NRM images, since phase and amplitude errors in the incoming
wavefront from a point source can be measured uniquely at each splodge.
Fig.~\ref{fig:pha} shows the phase of the complex visibility.
At $\bu = \bf{0}$ the phase of $\mathcal{V}$ is zero, since  $\mathcal{V}$ is a Hermitian function.

\section{The basics of image reconstruction} \label{sec:tech}
\emph{Correcting for pointing jitter:}
The variation of the pointing of a telescope during an exposure (jitter) leads
to image smearing,  as does the co-adding of short exposures. Image smearing
will limit the accuracy of deconvolution (Fig.~\ref{fig:naivecomp}, left
panels).  However, a pointing offset ($\Delta \bk$) generates a phase gradient
across $\mathcal{V(\bu}$, as predicted by the Fourier Shift theorem.  The
gradient is proportional to $|\Delta \bk|$, and is in the direction of $\Delta
\bk$.  For example, in Fig.~\ref{fig:pha} the phase of $\mathcal{V(\bu})$
slopes in the 8~o'clock to 2~o'clock direction.

Under ideal conditions, without a pointing offset from the centroid of the
image, the fringe phase is precisely the Fourier phase found by Fourier
transforming the image.  To align our images to sub-pixel accuracy (in order to
remove the effects of jitter), we fit a plane to the measured phase of
$\mathcal{V}(\bu)$,  making sure to constrain the constant term of the fit to
zero.  Subtracting the fitted phase gradient from the measured values of the
phase of $\mathcal{V}$ is completely equivalent to sub-pixel alignment of the
images.  Since we utilize only fringe information in subsequent data reduction,
we do not need to explicity reverse-transform our jitter-corrected
$\mathcal{V(}\bu)$ to co-align our image data frames.

In practice we perform a Fast Fourier Transform, padding our incoming image
data array by a factor of 4 to oversample $\mathcal{V}(\bu)$ for convenience
and clarity.  We fit the phase of the complex visibility within square boxes $N
\times N$ numerical pixels, centering the boxes on each splodge in the
oversampled $\bu$ plane.  This fit is relatively insensitive to our choice of
$N$, as long as $3<N < 41$.  We selected a value of $N=11$ for our data
reduction.  Including larger areas of the $\bu$ plane ($N>40$) in the fit
caused significant changes in the fit.  We presume this is because these larger
boxes mainly add noise to the fit.

\emph{Extraction:}
Ground-based optical, IR, and radio interferometry usually obtain a single
complex visibility measurement per baseline.  In the optical and IR, windowing
the image plane data numerically leads to an averaging (by convolution with the
Fourier transform of the windowing function) of $\mathcal{V}$ within a splodge.
In radio this is due to instrumental implementation, with one signal (ignoring
polariztion) measured by each antenna.  (We note that array feeds on radio
telescopes are beginning to change this paradigm).  With one measurement per
baseline, coverage in the $\bu$~plane is sparse. Fig.~\ref{fig:uvextract}a is
equivalent to a ground-based 7 aperture array.  In practice, baselines are moved by Earth's
rotation, which helps increase $\bu$~plane coverage over time.
Furthermore, the `zero-spacing' complex visibility, $\mathcal{V}({\bf 0})$ is
measured easily in these images.  This is in contrast to most (but not all)
radio interferometric measurements.
Because space-based optical and IR imaging is stable and repeatable, one can
obtain many more independent measurements of the complex visibility from a
single image.  This results in better instantaneous (u,v)~coverage.  We
tried a few different approaches to improving our imaging algorithms by
extracting more than one measurement per splodge.

\begin{figure*}
\centerline{
\includegraphics[angle=0,width=6.5in]{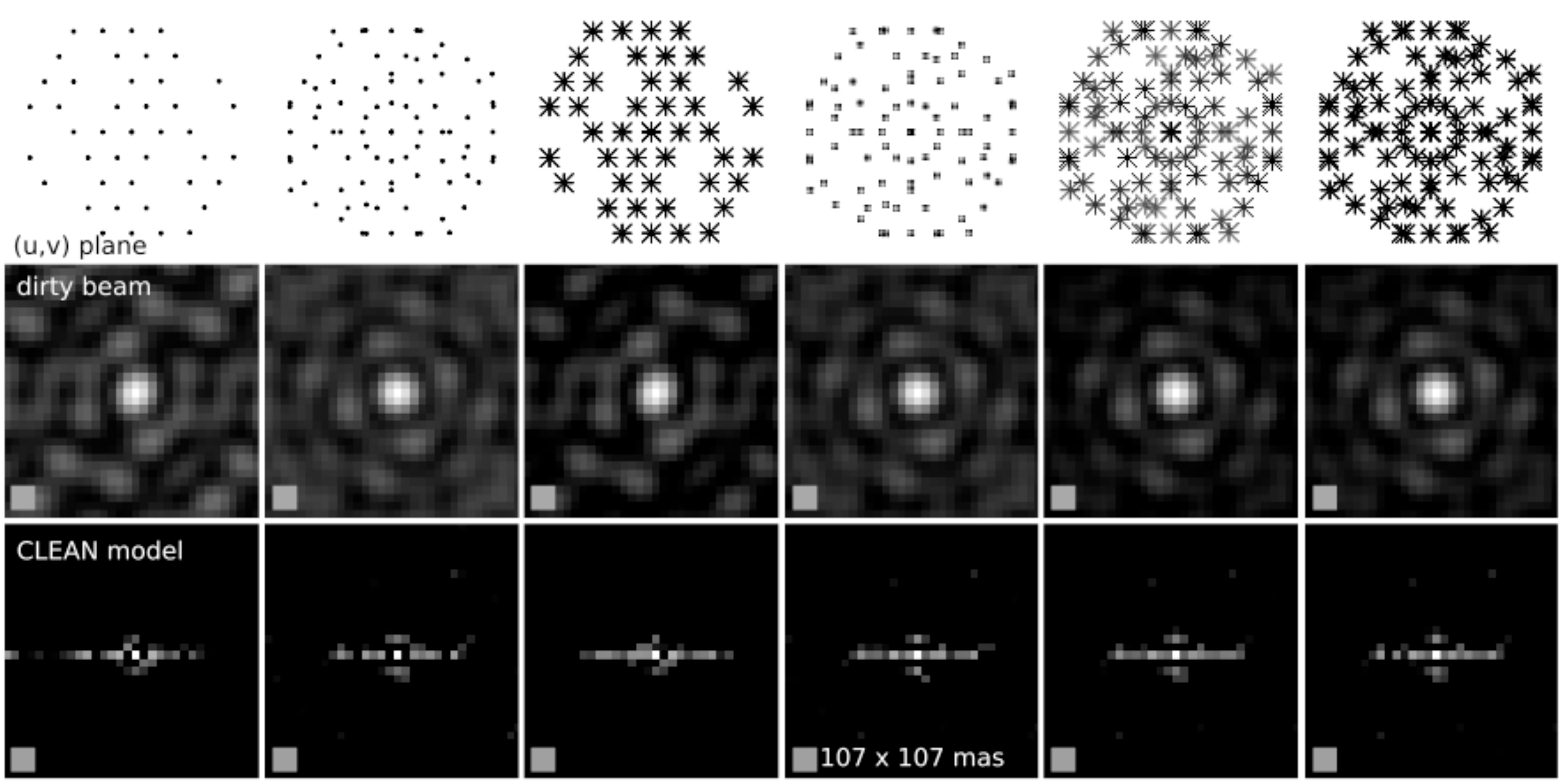}}
\caption{Effect of different extraction patterns on image reconstruction.
Each extraction pattern was used to reconstruct our fiducial model, shown
in fig~\ref{fig:tgt}a.
As the number of visibilities included increases, the number of artifacts decreases and the bar is more fully reconstructed.
\label{fig:clextract}}
\end{figure*}

Fig.~\ref{fig:clextract} shows how different extraction patterns change the PSF
and output \textsc{clean} model.  Extracting samples of $\mathcal{V}$ too
close to the edges of splodges adds more noise than signal (see
Fig.~\ref{fig:amp}).  Our standard extraction pattern is a $31 \times 31$ pixel box
centered on each splodge.  We read off values located at the bisectors and
vertices of each extraction box.
The optimal extraction pattern may be hexagonal, which would also allow simpler
combination with some types of weighting \citep[see][and below]{Briggs}.

\emph{Inversion:}
To inverse FT our phase corrected, extracted $\mathcal{V}$ and produce an image
($I^{\prime}_{\rm fringe}$--`dirty map') we choose: 1) a cell size $\delta$, i.e.
the size of a resolution element in the image and 2) the size of $I^{\prime}_{\rm fringe}$.
$\delta$ should be $\leq$ detector pixel scale (since information is available
on at least these scales).  We chose $\delta=0.25\lambda/D$ ($\sim 35.6$~mas). 
Reducing $\delta$ decreases the SNR per pixel, but optimal $\delta$
may be slightly smaller than our chosen value for some sources.  The size of
$I^{\prime}_{\rm fringe}$ must be large enough to prevent ringing, so we
typically choose to match the original detector image size (or $256\times 256$
pixels, for $\delta \sim 35.6$~mas)\footnote{For center-only extractions we must
make a smaller $I^{\prime}_{\rm fringe}$ of $128\times 128$ pixels to avoid
large sidelobes}.  Inverse FT potentially allows us to adjust the weighting of
$\mathcal{V}$ to e.g. highlight structures at particular length scales in $I^{\prime}_{\rm fringe}$
\citep{Briggs} by modulating amplitudes. For now we use natural weighting, but
will investigate optimal weighting for various sources in future work.

\emph{\textsc{Clean}ing and restoration:} \textsc{Clean} deconvolvers iteratively create a 
pixelated (gridded) model of the
the spatial distribution of sky brightness (in the region observed). The algorithms place a 
delta function (\textsc{clean} component) at the $\bk$ location
of the brightest pixel in the fringe image $I^{\prime}_{\rm fringe}$, convolve it with the PSF and subtract 
the result from $I^{\prime}_{\rm fringe}$; repeating with next 
brightest pixel until some condition is met (number of iterations, no negative flux, etc.). The `gain' parameter 
(we used gain=0.1) prevents oversubtracting 
on a single iteration. Because convolution and FT are linear
operations, the sum of many delta functions seems a useful approximation of the sky.  
However, (some radio interferometrists will object) the true sky is not discretized in this way \citep{Briggs}. 
On the other hand, our (NIRISS) detector \emph{is} pixelated, and so is our image of the sky. 
Therefore the \textsc{clean} model \emph{is} one proper comparator for us. 
Thus, prior models of a source (where available) should be compared to \textsc{clean} models, in
addition to the usual restored map.

We show \textsc{clean} models in conjunction with a restored map, $I_R=I^{\prime}_{\rm fringe}-(C*B)+(C*B_R)$ \citep{tcp99}. 
Here $*$ denotes convolution, $I^{\prime}_{\rm fringe}$ is the fringe image, $C$ is the clean model, 
$B$ is the PSF or `dirty beam' and $B_R$ is the `restoring beam'. 
We use a symmetric gaussian $B_R$ with a FWHM=71~mas ($0.5\lambda/D$). We also show a residual map, 
$R=I^{\prime}_{\rm fringe}-(C*B)$, in some cases. We can see that the restored map is just the \textsc{clean} model,
smoothed to our theoretical angular resolution, plus the residuals--as such, it provides a conservative
representation of the signifcance of reconstructed features in our images.

\section{Closure Phase \& Closure Amplitude}
\label{sec:cpca}
Closure phase (CP) and closure amplitude (CA) are methods allowing calibration and removal of 
aperture-specific errors and uncorrelated atmospheric noise \citep[e.g.][ and references therein]{MonnierAllen2012}. 
CP assumes that phase errors are dominated by noise uncorrelated between apertures (e.g. small-scale 
atmospheric distortions or antenna-specific noise). For any closed triangle of
apertures observing any source, such errors produce net zero \emph{displacement} of 
phase (because the vector sum is zero). 
The CP is:
\begin{equation}
\Phi_{ijk}=\phi_{ij}+\phi_{jk}+\phi_{ki}
\label{eq:cp}
\end{equation}
where $\phi_{ij}$ is the phase measured along the baseline between apertures $i$ and $j$, etc. (and $\Phi_{ijk}=0$ for a point source).  
Given $N$ apertures, there are $N \choose 3$ closed triangles, but only ${N-1} \choose 2$ of these are 
independent (with $N \choose 2$ independent phases). CP is insensitive to phase errors 
which \emph{are} correlated across apertures (which does occur in the atmosphere). Correlated errors cannot be calibrated out by CP; 
this is why ground-based O/IR interferometry cannot measure fringe phase.

CA is a similar technique, utilizing independent quads of apertures \citep[e.g.][]{readhead80}. CA uses ratios of amplitudes along pairs
of baselines, i.e. for apertures $i, j, k, l$ the CA is:
\begin{equation}
\Gamma_{ijkl}=\frac{A^{\prime}_{ij}A^{\prime}_{kl}} {A^{\prime}_{ik}A^{\prime}_{jl}}
\label{eq:ca}
\end{equation}
where $A^{\prime}_{ij}$ is the amplitude measured along the baseline between apertures $i$ and $j$, etc.
There are $N(N-3)/2$ independent closure amplitudes.
As long as $A_{ij}=g_ig^*_jA^{\prime}_{ij}$, where $A_{ij}$ is the true amplitude, $A^{\prime}_{ij}$ is the measured amplitude
and $g_i$ and $g^*_j$ are the (uncorrelated) complex aperture specific gain and conjugate (and dominate the amplitude errors),
the measured and true CAs are equal (and $\Gamma_{ijkl}=1$ for a point source). However, amplitudes in 
ground-based O/IR interferometry are dominated by correlated noise from the atmosphere so CA cannot be used \citep[e.g.][]{MonnierAllen2012}.
Note also that the gain is an average over the entire aperture; when extracting $A^{\prime}$ away from the
center of a splodge, we are effectively creating many subapertures. The gain for such
subapertures is not clearly defined and does not have a simple relationship to the gain defined over the whole
aperture. This is why it is mathematically incorrect to apply CP and CA relationships to visibilities extracted away from the splodge centers.

\bibliographystyle{apj}
\bibliography{nrm_agn_bibtex.bib}{}


\label{lastpage}

\end{document}